\documentclass[5p,twocolumn]{elsarticle} 

\journal{Automatica}
\biboptions{numbers,sort&compress}

\newproof{proof}{Proof}

\usepackage{natbib}
                               
\usepackage{graphicx,fleqn,amstext,amsmath,amssymb,color,psfrag,mathabx,booktabs}
\usepackage[utf8]{inputenc}
\usepackage[colorlinks=true, letterpaper,    
urlcolor=black,     
citecolor=blue,
menucolor=black,    
linkcolor=black,    
pagecolor=black,    
breaklinks=true,
]{hyperref}

\newcommand{\R}{\ensuremath{\mathbb{R}}}

\newcommand{\Rlo}{\ensuremath{\mathbb{R}_{\geq 0}}}

\newcommand{\Zo}{\ensuremath{\mathbb{Z}_{\geq 0}}}
\newcommand{\Zp}{\ensuremath{\mathbb{Z}_{> 0}}}
\newcommand{\N}{\ensuremath{\mathbb{N}}}                 
\newcommand{\Z}{\ensuremath{\mathbb{Z}}}

\definecolor{bleucit}{rgb}{0.2,0.4,0.6} 

\definecolor{blue_cv}{rgb}{0.09,0.35,0.78}

\newcommand{\KL}{\ensuremath{\mathcal{KL}}}
\newcommand{\K}{\ensuremath{\mathcal{K}}}
\newcommand{\Kinf}{\ensuremath{\mathcal{K}_{\infty}}}

\newcommand{\dom}{\ensuremath{\text{dom}\,}}

\newcommand{\norm}[1]{\ensuremath{\left\|{#1}\right\|}}

\newtheorem{ass}{\textnormal{\textbf{Assumption}}}
\newtheorem{prop}{Proposition}

\newtheorem{lem}{Lemma}

\newtheorem{thm}{\textnormal{\textbf{Theorem}}}

\newtheorem{rem}{\textnormal{\textbf{Remark}}}

\usepackage{mathtools}

\usepackage{enumitem}

\usepackage{tikz}
\usetikzlibrary{shapes,arrows,calc}
\definecolor{MyGreen}{RGB}{50,140,80}

\begin{document}

\begin{frontmatter}
                                              

\title{\textbf{Decentralized event-triggered estimation of nonlinear systems}} 


\tnotetext[t1]{This work was funded by Lorraine Universit\'e d'Excellence LUE, HANDY project ANR-18-CE40-0010-02, the France Australian collaboration project IRP-ARS CNRS and the Australian Research Council under the Discovery Project DP200101303.}

\author[Nancy]{E. Petri}\ead{elena.petri@univ-lorraine.fr} 
\author[Nancy]{R. Postoyan}\ead{romain.postoyan@univ-lorraine.fr}   
\author[Lyon]{D. Astolfi}\ead{daniele.astolfi@univ-lyon1.fr}               
\author[Melbourne]{D. Ne\v{s}i\'c}\ead{dnesic@unimelb.edu.au}  
\author[Eindhoven]{W.P.M.H. Heemels}\ead{m.heemels@tue.nl}

\address[Nancy]{Universit\'e de Lorraine, CNRS, CRAN, F-54000 Nancy, France. }  
\address[Lyon]{Universit\'e Claude Bernard Lyon 1, CNRS, LAGEPP UMR 5007, F-69100,
	Villeurbanne, France.}  
\address[Melbourne]{Department of Electrical and Electronic Engineering, The University of Melbourne, Parkville, 3010 Victoria, Australia.}        
\address[Eindhoven]{Department of Mechanical Engineering,
	Eindhoven University of Technology, The Netherlands.}  
          

\begin{abstract}                          
We investigate the scenario where a perturbed nonlinear system transmits its output measurements to a remote observer via a packet-based communication network. The sensors are grouped into $N$ nodes and each of these nodes decides when its measured data is transmitted over the network independently. The objective is to design both the observer and the local transmission policies in order to obtain accurate state estimates, while only sporadically using the communication network. 
In particular, given a general nonlinear observer designed in continuous-time satisfying an input-to-state stability property, we explain how to systematically design a dynamic event-triggering rule for each sensor node that avoids the use of a copy of the observer, thereby keeping local calculation simple.
 We prove the practical convergence property of the estimation error to the origin and 
 we show that there exists a uniform strictly positive minimum inter-event time for each local triggering rule under mild conditions on the plant. The efficiency of the proposed techniques is illustrated on a numerical case study of a flexible robotic arm.
\end{abstract}

\end{frontmatter}

\section{Introduction}
While digital networks exhibit a range of benefits for control applications in terms of ease of installation, maintenance and reduced weight and volume, they also require adapted control theoretical tools to cope with the induced communication constraints (e.g., sampling, delays, packet drops, scheduling, quantization), see e.g., \cite{hespanha2007survey, heemels2010stability}. In this work, we concentrate on the state estimation of nonlinear systems over a digital channel and we focus on the effect of sampling. In particular, we consider 
state estimation where the plant is nonlinear, perturbed and communicates its measurements over a digital network to a remote observer, whose goal is to estimate the plant state.
The communication schedule is very important to guarantee good estimation performance. 
An option is to generate the transmission instants based on time, in which case we talk of time-triggered strategies for which various results are available in the literature, see, e.g., \cite{postoyan2011framework, li2017robust, ferrante2016state, mazenc2015design, davcic2008observer}. However, this paradigm may generate (significantly) more transmissions over the network than necessary to fulfill the estimation task, thereby leading to a waste of the network resources. 
 As a potential and promising solution, one can use event-triggered transmissions to overcome this drawback, see e.g., \cite{heemels2012introduction} and the references therein. 
In this case, an event-based triggering rule monitors the plant measurement and/or the observer state and decides when an output transmission is needed. 

Various event-triggered techniques are available in the literature for estimation, see, e.g., \cite{scheres2021Event, li2010event,shi2014event2,li2011performance,trimpe2014stability, yu2021stochastic, song2021event, shi2016event, huang2019robust, sijs2012event, hu2020event, etienne2017periodic,etienne2016event, etienne2017asynchronous, tong2020finite, niu2020dynamic}. 
Numerous papers propose to implement a copy of the observer within the sensor and then use its information to define the transmission instants, 
see e.g., \cite{scheres2021Event, li2010event,shi2014event2,li2011performance,trimpe2014stability, yu2021stochastic, song2021event}.
A possible drawback with this technique is that it may require significant computation capabilities on the sensors, especially in the case of large-scale systems, or highly nonlinear dynamics, which may be unavailable.
Another solution is to follow an event-triggered strategy, which is only based on  a static condition involving the measured output and its past transmitted value(s) see, e.g., \cite{shi2016event, huang2019robust, sijs2012event, hu2020event, etienne2017periodic,etienne2016event, etienne2017asynchronous, tong2020finite}. Consequently, it is not necessary to implement a copy of the observer in the sensors and thus the sensors are not required to have significant computation capabilities. 
However, such static triggering rules may generate a lot of transmissions and the results in \cite{shi2016event, huang2019robust, sijs2012event, hu2020event, etienne2017periodic,etienne2016event, etienne2017asynchronous, tong2020finite} only apply to specific classes of systems and 
a centralized scenario, where all sensors communicate simultaneously over the network, with the exception of \cite{shi2016event, hu2020event}. 
An alternative are self-triggering policies, see e.g., \cite{andrieu2015self, rabehi2020finite}, where the observer requests a new output measurement when it needs it to perform the estimation. However, the available results only apply to specific classes of systems. Moreover, self-triggering rules typically generate more transmissions than event-triggered ones. 

In this paper, we adopt a dynamic event-triggered approach based only on the measured output and the last transmitted output value. This strategy keeps monitoring
 the plant output, and thereby may lead to less transmissions compared to a self-triggering approach. Moreover, it does not require a copy of the observer, which simplifies the implementation and requires less computation capability on the sensor. 
The main novelties are, first, the design of a new triggering rule, which involves an auxiliary scalar variable for each sensor node, that has several benefits as explained in the sequel. 
Second, the proposed results apply to general, perturbed nonlinear systems contrary to the vast majority of works in the literature, which concentrates on specific classes of systems, see e.g., \cite{li2010event,shi2014event2,li2011performance,trimpe2014stability, yu2021stochastic, song2021event, shi2016event, huang2019robust, sijs2012event, hu2020event, etienne2017periodic,etienne2016event, etienne2017asynchronous, tong2020finite, niu2020dynamic}. Third, the triggering strategies are decentralized.  Indeed, we consider the scenario with 
 $N$ sensor nodes, where each node decides independently when to transmit its local data to the observer via a digital network. Consequently, each sensor node has its own triggering rule.

Our design is following an emulation-based approach in the sense that in the first step the observer is designed ignoring the effects of the communication network. 
In particular, we assume that the observer has been synthesized in continuous-time in such a way that it satisfies an input-to-state stability property, that holds for many observer design techniques of the literature, see e.g., \cite{astolfi2021stubborn, shim2015nonlinear} and the references therein.
In the second step, we take the network into account and propose a new hybrid model using the formalism of \cite{goedel2012hybrid1, heemels2021hybrid}. We then design a dynamic triggering rule for each sensor node to approximately preserve the original properties of the observer. In particular, we ensure that the estimation error system satisfies a global practical stability property  
and we show that, in some particular cases, it is possible to recover the same decay rate for the Lyapunov function along the solutions as in the absence of the communication network. Note that, we do not guarantee an asymptotic stability property, but a practical one in general, which is a consequence of the absence of a copy of 
the observer in the triggering mechanism as we explain later (see Remark~\ref{RemarkWhyOnlyPractical}).
%
As already stated, the triggering rules are dynamic in the sense that they involve a local scalar auxiliary variable, which essentially filters an absolute threshold type condition, see e.g., \cite{etienne2017periodic,etienne2017asynchronous,etienne2016event,  tong2020finite}. This is a new in the context of estimation, to the best of the authors' knowledge, and is inspired by related event-triggering control techniques \cite{girard2014dynamic, tanwani2015using, tabuada2007event}. 
In addition, our design of the triggering rules rely on very mild knowledge of the observer properties; only some qualitative knowledge is needed on the gains appearing in the input-to-state stability dissipativity property, which is assumed to hold for the state estimation error system, as will be explained in more detail below. 

Compared to \cite{shi2016event, huang2019robust, sijs2012event, hu2020event}, we do not consider a stochastic setting and discrete-time plants, but deterministic (nonlinear) continuous-time systems, which raise the issue of potential Zeno phenomena. 
\color{black}Moreover, in our work we propose a new triggering rule, which filters the absolute threshold rule proposed in e.g., \cite{etienne2017periodic,etienne2016event, etienne2017asynchronous, tong2020finite} and, as a result, typically leads to less transmissions, as illustrated on a numerical robot example in this paper.  \color{black}
	The closest work is \cite{niu2020dynamic} where a similar triggering rule is presented, but only for polynomial systems and for a centralized approach (one communication sensor node only). In contrary, our results essentially only rely on an input-to-state stability assumption of the estimation error system, which is commonly satisfied \cite{astolfi2021stubborn}.
	Moreover we consider the more challenging case of a decentralized set-up, we provide in-depth characterizations of the domains of the solutions and we provide various extensions for scenarios where the outputs are affected by additive noise, and where the plant input is also transmitted over the network (see Section~\ref{discussionSection}).
Compared to our preliminary version of this work \cite{petri2021Event}, here we consider nonlinear systems, instead of linear time-invariant ones, and the transmission strategy is decentralized, and not centralized as in \cite{petri2021Event}. 
Moreover, the plant is affected by unknown disturbances and we prove the completeness of maximal solutions for the overall system.

The remainder of the paper is organized as follows. Preliminaries are stated in Section~\ref{Notation}. The problem setting, the assumption on the observer and the problem statement are presented in Section~\ref{ProblemStatement}. The proposed triggering rule and the overall hybrid system model are given in Section~\ref{TriggeringRuleAndHybridModel}. In Section~\ref{StabilityGuarantees} we analyze the stability properties of the proposed event-triggered observer. In Section~\ref{SolutionDomainSection} we derive various properties of the solutions domains (completeness of maximal solutions, existence of  a minimum time between any two transmissions of each sensor node as well as a condition that allows transmissions to stop). 
Some generalizations and extensions are presented in Section~\ref{discussionSection} and a numerical case study on a flexible joint robotic arm is reported in  Section~\ref{Example}. Finally, Section~\ref{Conclusions} concludes the paper. Two technical lemmas are given in the Appendix.
\section{Preliminaries}\label{Notation} 
The notation $\R$ stands for the set of real numbers and $\Rlo:= [0, +\infty)$. 
We use $\Z$ to denote the set of integers, $\Zo:= \{0,1,2,...\}$ and $\Zp:= \{1,2,...\}$. For a vector $x \in \R^n$, $|x|$ denotes its Euclidean norm. For a matrix $A \in \R^{n  \times m}, \norm{A}$ stands for its 2-induced norm. For any signal $v: \R_{\geq0} \to \R^{n_v}$, with $n_v \in \Zp$, $\norm{v}_{[t_1, t_2]}:= \textnormal{ess} \sup_{t \in [t_1, t_2]} |v(t)|$. Given a real, symmetric matrix $P$, its maximum (minimum) eigenvalue is denoted $\lambda_{\max}(P) \  (\lambda_{\min}(P))$.  The notation $I_N$ stands for the identity matrix of dimension $N \in \Zp$, while $0_{N \times M}$ stands for the null matrix of dimension $N \times M$, with $N, M \in \Zp$. We consider class-$\K$, $\Kinf$, $\KL$ functions as defined in \cite{goedel2012hybrid1}. 
We model hybrid systems in the formalism of \cite{goedel2012hybrid1, heemels2021hybrid}, namely
\begin{equation}
	\mathcal{H} \;:\; \left\{
	\begin{array}{rcll}
		\dot x &=& F(x,u), & \quad (x,u)\in \mathcal{C}, 
		\\
		x^+ & \in & G(x,u),  &\quad (x,u)\in \mathcal{D},
	\end{array}
	\right.
\end{equation}
where 
$\mathcal{C}\subseteq \R^{n_x} \times \R^{n_u} $ is the flow set, 
$\mathcal{D}\subseteq \R^{n_x} \times \R^{n_u}$ is the jump set,
$F$ is the flow map and $G$ is the jump map. 
Solutions  
to system (1) are defined on  \textit{hybrid time domains}. A set 
$E\subset \Rlo\times \Zo$ is a \textit{compact 
	hybrid time domain} if $E = \bigcup_{j=0}^{J-1}([t_j, t_{j+1}], j)$
for some finite sequence of times $0=t_0\leq t_1\leq \ldots \leq t_{J}$
and it is a \textit{hybrid time domain} if for all $(T,J)\in E$, 
$E\cap ([0,T]\times \{0,1,\ldots, J\})$
is a compact hybrid time domain. Given a hybrid time domain $E$, $
\sup_{j}E := \sup \{j\in \Zo: \exists \, t\in \Rlo \textnormal{ such that } (t,j)\in E
\}$.
A hybrid signal $x : \dom \,x \,\to \R^{n_x}$ is called a \textit{hybrid arc} 
if $x(\cdot, j)$ is locally absolutely continuous for each $j$. 
Given a set $\mathcal{U} \subseteq \R^{n_u}$, $\mathcal{L}_{\mathcal{U}}$ is the set of all functions from $\R_{\geq0}$ to $\mathcal{U}$ that are Lebesgue measurable and locally essentially bounded. 
We consider the notion of solution proposed in \cite{heemels2021hybrid}. Hence, a hybrid arc $x$ 
is a \textit{solution} to $\mathcal{H}$ for a given input $u \in \mathcal{L}_{\mathcal{U}}$, 
if 
\begin{itemize}
	\item for all $j\in \N$ such that $I^{j}:= \{t \ | \ (t,j) \in \dom \, x \}$ has nonempty interior,  $\dot{x}(t,j) \in F(x(t,j), u(t))$  and $(x(t,j), u(t)) \in \mathcal{C}$ for almost all $t \in I^{j}$; 
	\item for all $(t,j) \in \dom \, x$ such that $(t, j+1) \in \dom \, x$, $(x(t,j), u(t)) \in \mathcal{D}$ and $x(t,j+1) \in \mathcal{G}(x(t,j), u(t))$.
\end{itemize}
%
A solution $x$ to $\mathcal{H}$  for a given input $u \in \mathcal{L}_\mathcal{U}$ is \textit{maximal}, if there does not exist another solution $\tilde{x}$ to $\mathcal{H}$ for the same input $u$ such that $\dom \, x$ is a proper subset of $\dom \, \tilde{x}$ and $x(t,j) = \tilde{x} (t,j)$ for all $(t,j) \in \dom \, x$. Moreover, a maximal solution $x$ to $\mathcal{H}$  for a given input $u \in \mathcal{L}_\mathcal{U}$ is \textit{complete}, if $\dom x$ is unbounded.

\section{Problem statement}\label{ProblemStatement}
\subsection{Setting}
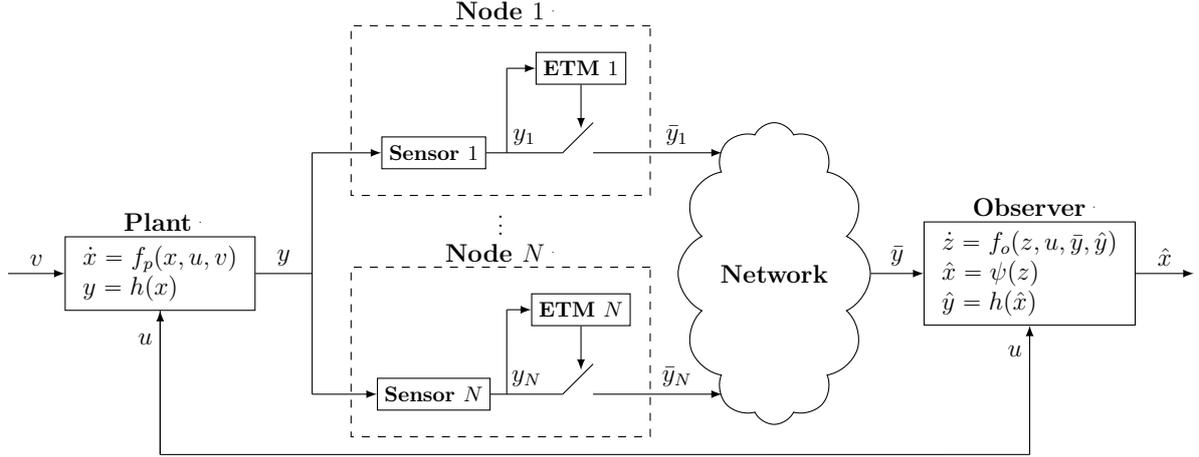
\begin{figure*}[]
	\begin{center}
		\tikzstyle{blockB} = [draw, fill=blue!30, rectangle, 
		minimum height=2em, minimum width=3em]  
		\tikzstyle{blockG} = [draw, fill=MyGreen!40, rectangle, 
		minimum height=2em, minimum width=3em]
		\tikzstyle{blockR} = [draw, fill=red!40, rectangle, 
		minimum height=2em, minimum width=3em]
		\tikzstyle{blockO} = [draw,minimum height=1.5em, fill=orange!20, minimum width=4em]
		\tikzstyle{input} = [coordinate]
		\tikzstyle{blockW} = [draw,minimum height=1.5em, fill=white!20, minimum width=4em]
		\tikzstyle{input1} = [coordinate]
		\tikzstyle{blockCircle} = [draw, circle]
		\tikzstyle{sum} = [draw, circle, minimum size=.3cm]
		\tikzstyle{blockSensor} = [draw, fill=white!20, draw= blue!80, line width= 0.8mm, minimum height=10em, minimum width=13em]
		\tikzstyle{blockDOT} = [draw,minimum height=8em, fill=white!20, minimum width=14em, dashed]
		
		\begin{tikzpicture}[auto, node distance=2cm,>=latex , scale=0.80,transform shape] 
			
			\node [input, name=input] {};
			\node [blockW, above of=input, node distance=3cm] (plant) {   
				\large 
				$		\begin{array}{l}
					\dot{x} = f_p(x,u,v)\\ y = h(x)
				\end{array}$};
			\node [input, above of= plant, node distance=0.85cm, right = 0.65 cm] (plantName) {};
			\draw [] (plantName) -- node  [pos=0.5]{\textbf{\large Plant}} (plantName);
			
			\draw [draw,->] (input) -- node {} (plant);
			\draw [draw,-] (input) -- node [pos=0.8]{\large $u$} (plant);
			
			\node [input, left of= plant, node distance=2.5cm] (DisturbanceInput) {};
			\draw [draw,->] (DisturbanceInput) -- node [pos=0.5]{\large $v$} (plant);

			\node [input, right of= plant, node distance=2.5cm] (triggering) {};
			\draw [draw,-] (plant) -- node  [pos=0.5]{\large $y$} (triggering);
			\node [input, above of= triggering, node distance=2cm] (sensor1line) {};
			\node [input, above of= sensor1line, node distance=0.7cm] (node1high) {};
			\node [blockDOT, right of=node1high, node distance=3.1cm] (node1) {};
			\node [input, above of= node1, node distance=1.65cm, right = 0.85 cm] (node1Name) {};
			\draw [] (node1Name) -- node  [pos=0.5]{\textbf{\large Node $1$}} (node1Name);
			
			\draw [draw,-] (triggering) -- (sensor1line);
			\node [input, below of= triggering, node distance=2cm] (sensorNline) {}; 
			
			\node [input, above of= sensorNline, node distance=0.7cm] (nodeNhigh) {};
			\node [blockDOT, right of=nodeNhigh, node distance=3.1cm] (nodeN) {};
			\node [input, above of= nodeN, node distance=1.65cm, right = 0.85 cm] (nodeNName) {};
			\draw [] (nodeNName) -- node  [pos=0.5]{\textbf{\large Node $N$}} (nodeNName);

			\draw [draw,-] (triggering) -- (sensorNline);
			\node [blockW, right of=sensor1line, node distance=2cm] (sensor1) {
				\textbf{Sensor $1$}
			};
			\draw [draw,->] (sensor1line) -- (sensor1);
			\node [blockW, right of=sensorNline, node distance=2cm] (sensorN) {
				\textbf{Sensor $N$}
			};
			\draw [draw,->] (sensorNline) -- (sensorN);
			\node at ($(node1)!.45!(nodeN)$) {\vdots};
			
			\node [input, right of= sensor1, node distance=2.12cm] (sampling1) {};
			\node [input, right of= sampling1, node distance=0.5cm] (sampling1right) {};
			\node [input, above of= sampling1right, node distance=0.5cm] (sampling1final) {};
			\draw [draw,-] (sampling1) -- (sampling1final);
			\draw [draw,-] (sensor1) -- node  [pos=0.5]{\large $y_1$}(sampling1);
			\node [input, right of= sensorN, node distance=2.12cm] (samplingN) {};
			\node [input, right of= samplingN, node distance=0.5cm] (samplingNright) {};
			\node [input, above of= samplingNright, node distance=0.5cm] (samplingNfinal) {};
			\draw [draw,-] (samplingN) -- (samplingNfinal);
			\draw [draw,-] (sensorN) -- node  [pos=0.5]{\large $y_N$} (samplingN);
			
			\node [input, above of= sampling1, node distance=1.4cm] (ETM1temp) {};
			\node [blockW, right of=ETM1temp, node distance=0.3cm] (ETM1) {
				\textbf{ETM $1$}
			};
			\node [input, right of= sensor1, node distance=1.2cm] (ETM1input) {};
			\node [input, above of= ETM1input, node distance=1.4cm] (ETM1inputBis) {};
			\draw [draw,-] (ETM1input) -- (ETM1inputBis);
			\draw [draw,->] (ETM1inputBis) --  (ETM1);
			\node [input, below of= ETM1, node distance=1cm] (ETM1arrow) {};
			\draw [draw,->] (ETM1) --  (ETM1arrow);
			\node [input, above of= samplingN, node distance=1.4cm] (ETMNtemp) {};
			\node [blockW, right of=ETMNtemp, node distance=0.3cm] (ETMN) {
				\textbf{ETM $N$}
			};
			\node [input, right of= sensorN, node distance=1.2cm] (ETMNinput) {};
			\node [input, above of= ETMNinput, node distance=1.4cm] (ETMNinputBis) {};
			\draw [draw,-] (ETMNinput) -- (ETMNinputBis);
			\draw [draw,->] (ETMNinputBis) --  (ETMN);
			\node [input, below of= ETMN, node distance=1cm] (ETMNarrow) {};
			\draw [draw,->] (ETMN) --  (ETMNarrow);

			\node [cloud, draw,cloud puffs=12,cloud puff arc=130, aspect=2, inner ysep=1.1em, right of=triggering, node distance=7.6cm, minimum height=5cm,
			fill =white!30] (network){\textbf{\large Network}};
			\node [input, right of= sampling1right, node distance=2.12cm] (network1) {};
			\draw [draw,->] (sampling1right) -- node [pos=0.65]{\large $\bar{y}_1$} (network1);
			\node [input, right of= samplingNright, node distance=2.12cm] (networkN) {};
			\draw [draw,->] (samplingNright) -- node [pos=0.65]{\large $\bar{y}_N$} (networkN);
			
			\node [blockW, right of=network, node distance=4.2cm] (observer) {
				\large 
				$\begin{array}{l}
					\dot{z} = f_o(z,u,\bar{y},\hat{y})\\
					\hat{x} = \psi(z)\\
					\hat{y} = h(\hat{x})\end{array}	$
			};
			\node [input, above of= observer, node distance=1.1cm, right = 1.05 cm] (observerName){};
			\draw [] (observerName) -- node  [pos=0.5]{\textbf{\large Observer}} (observerName);
			\draw [draw,->] (network) -- node [pos=0.5]{\large $\bar{y}$} (observer);
			\node [input, below of=observer, node distance=3cm] (input2) {};
			\draw [draw,->] (input2) -- node [pos=0.8]{\large $u$} (observer);
			\node [input, right of=observer, node distance=2.7cm] (stateEstimate) {};
			\draw [draw,->] (observer) -- node [pos=0.5]{\large $\hat{x}$} (stateEstimate);
			\draw [draw,-] (input) -- (input2);
			\node [input, right of= input, node distance=6.3cm] (input3){};
			\node [input, below of= input3, node distance=0.4cm] (input4){};
			
		\end{tikzpicture}
	\end{center}
	\caption{Block diagram representing the system architecture (ETM: Event-Triggering Mechanism)}
	\label{Fig:blockDiagramDistributed}
\end{figure*}
%
Consider the nonlinear system
\begin{equation}
		\dot x  =  f_p(x,u,v),  \qquad
		y  =  h(x) ,
	\label{eq:system}
\end{equation}
where $x(t) \in \R^{n_x}$ is the state to be estimated by the observer, $u(t) \in \R^{n_u}$ is the measured input, $y(t) \in \R^{n_y}$ is the output measured by sensors, and $v(t) \in \R^{n_v}$ is an unmeasured disturbance input 
at time $t \in \R_{\geq0}$
with $n_x, n_y$ $\in \Zp$, and $n_u , n_v \in \Zo$. The inputs $u$ and $v$ to \eqref{eq:system} are such that $u \in \mathcal{L}_{\mathcal{U}}$ and $v \in \mathcal{L}_{\mathcal{V}}$ for some sets $\mathcal{U} \subseteq \R^{n_u}$ and $\mathcal{V} \subseteq \R^{n_v}$. 
The vector field $f_p: \R^{n_x} \times \R^{n_u} \times \R^{n_v} \rightarrow \R^{n_x}$ is locally Lipschitz in its first argument and continuous in the others and $h: \R^{n_x} \rightarrow \R^{n_y}$ is continuously differentiable. 

We follow an emulation-based design in the sense that a continuous-time observer for system \eqref{eq:system} is first designed ignoring the packet-based nature of the communication network.
Afterwards, we will consider the network and design a triggering rule to decide when the output data need to be transmitted to the observer in order to approximately preserve its original properties.
In particular, we assume the availability of a continuous-time observer for system \eqref{eq:system} of the form
\begin{equation}
	\begin{aligned}
		\dot{z}  &= f_o(z,u,y,\hat{y}), \\
		\hat{x} &= \psi(z), \qquad 
		\hat y  = h(\hat{x}),
	\end{aligned}
	\label{eq:observerAlly}
\end{equation}
where $z(t) \in \R^{n_z}$ is the observer state, with $n_z \geq n_x$, $\hat{x}(t) \in \R^{n_x}$ is the state estimate, $\hat{y} (t)$ 
is the output estimate at time $t \in \R_{\geq0}$. The vector field $f_o: \R^{n_z} \times \R^{n_u} \times \R^{n_y} \times \R^{n_y} \rightarrow \R^{n_z}$ is continuous, and $\psi: \R^{n_z} \rightarrow \R^{n_x}$ admits a right inverse $\psi^{-R}$ of $\psi$, i.e., $x = \psi(\psi^{-R}(x))$ for any $x \in \R^{n_x}$. 
Often $z = \hat{x}$, but this does not necessary have to be the case, like in Kalman filters, which involve extra variables that can be stacked in vector $z$.
Observer~\eqref{eq:observerAlly} has a general structure and can be designed using several 
observer design procedures, including Luenberger-like observers and Kalman filters, see e.g., \cite{astolfi2021stubborn, postoyan2011framework}, \cite[Section IV]{shim2015nonlinear} and the references therein.
The precise assumption we make on observer \eqref{eq:observerAlly} is stated later in this section.
	For simplicity, we do not consider in this work the case of reduced-order observers (see e.g., \cite{postoyan2012emulated}), but we believe that similar derivations could be developed in this scenario. 
We also adopt the following assumption.
\begin{ass}\itshape
		The observer has access to the input $u$ at any time instant. \hfill $\Box$ 
		\label{assumption 1}
	\end{ass}
	Assumption~\ref{assumption 1}  is reasonable in many control applications such as, for example, when the control input is generated on the observer side. 
	It is worth noting that, when the observer does not know the input $u$, meaning that Assumption~\ref{assumption 1} is not satisfied, the input $u$ can be included in the unknown disturbance input $v$ in \eqref{eq:system} and the results presented in the sequel apply, as long as Assumption~\ref{ISSassumption} presented later holds. Furthermore, in the case where the input $u$ is transmitted from the plant to the observer via a digital network, we explain in Section~\ref{TriggeredInput} how to define a triggering rule for $u$ so that the forthcoming results hold \textit{mutatis mutandis}.

We investigate the scenario where the output measurements of system~\eqref{eq:system} are transmitted to observer~\eqref{eq:observerAlly} via a digital channel, as depicted in Figure~\ref{Fig:blockDiagramDistributed}. 
In particular, we consider the setup where the sensors are grouped into  $N$ nodes, where $N \in \{1, \dots, n_y\}$ and we write, after re-ordering (if necessary), $y = (y_1, \dots , y_N) = (h_1(x), \dots, h_N(x))$ with $y_i \in \R^{n_{y_i}}$, $n_{y_i} \in \{1, \dots, n_y\}$ and $n_{y{_1}} + \ldots + n_{y_N} = n_y$.
Each sensor node decides when its output measurement needs to be transmitted to the observer over the network, independently of the other sensor nodes. 
Hence several nodes are allowed to communicate at the same time instant. 

In this setting, the observer does not know $y$, but its networked version $\bar y := (\bar{y}_1, \dots , \bar{y}_N) \in \R^{n_y}$. Each $\bar{y}_i \in \R^{n_{y_i}}$, with $i \in \{1,\dots, N\}$, is generated with a zero-order-hold device between two successive transmission instants, i.e., in terms of the hybrid systems notation of Section~\ref{Notation},
\begin{equation}
	\dot{\bar{y}}_i = 0 \\
	\label{eq:bar_yi}
\end{equation}
and, when a transmission of node $i$ occurs the corresponding output $y_i$ is transmitted, considering an ideal sampler, hence
\begin{equation}
	\bar{y}_i^+ = y_i,
	\label{eq:yplus}
\end{equation}
otherwise, when another node generates a transmission the last received value is kept constant, i.e.
\begin{equation}
	\bar{y}_i^+ = \bar{y}_i.
	\label{eq:yplusBis}
\end{equation}
It is worth noting that the zero-order-hold is just a choice we make to generate the output sampled version $\bar{y}_i$ for all $i \in \{1, \dots, N\}$ between transmission times. Other options are for example the first-order-hold and the model-based holding function \cite{lunze2010state}. 

Since the output $y$ is transmitted over the network, observer \eqref{eq:observerAlly} 
does not have access to the exact measurement output $y$, but its networked version $\bar{y}$.
As a result, the observer equations in \eqref{eq:observerAlly} become
\begin{equation}
	\begin{aligned}
		\dot{z}  &= f_o(z,u,\bar{y},\hat{y}), \\
		\hat{x} &= \psi(z), \qquad
		\hat y  = h(\hat{x}).
	\end{aligned}
	\label{eq:observerNew}
\end{equation}
We define the network-induced error for each sensor node $e_i:= \bar {y}_i - y_i  \in \R^{n_{y_i}}$, with $i \in \{1,\dots, N\}$, and the concatenated vector $e:= (e_1, \dots, e_N) = \bar y - y  \in \R^{n_y}$. We obtain, in view of \eqref{eq:system} and \eqref{eq:observerNew},
\begin{equation}
	\dot{z}  = f_o(z,u, y + e,\hat{y}) = f_o(z,u, h(x) + e, h(\psi(z))). \\
	\label{eq:observer}
\end{equation}
The dynamics of variable $e_i$, for $i \in \{1, \dots, N\}$, between two successive transmission instants is, in view of \eqref{eq:system} and \eqref{eq:bar_yi} and since $h_i$ is (continuously) differentiable,  
\begin{equation}
	\dot{e}_i  = \dot{\bar{y}}_i - \dot{y}_i  = -\frac{\partial h_i(x)}{\partial x}f_p(x,u,v) =: g_i(x,u,v).\\
	\label{eq:samplingErrorContinuous}
\end{equation}
Furthermore, at each transmission instant of the $i$-th sensor node, we have 
\begin{equation}
		e_i^+ = 0,
	\label{eq:samplingErrorDiscrete}
\end{equation}
in view of \eqref{eq:yplus}, while, for $j \in \{1, \dots, N\}$ with $j \neq i$,
\begin{equation}
	e_j^+ = e_j.
	\label{eq:samplingErrorDiscreteBis}
\end{equation}
\subsection{Assumption on the observer}
Inspired by \cite{astolfi2021stubborn}, we require observer \eqref{eq:observerAlly} to satisfy the next input-to-state stability property. 
\begin{ass}
	There exist $\underline{\alpha}$, $\overline{\alpha}$, $\alpha$, $\gamma_1, \dots, \gamma_N$$, \theta$ $\in \Kinf$, $V: \R^{n_x} \times \R^{n_z} \rightarrow \R_{\geq0}$ continuously differentiable, such that for all $x \in \R^{n_x}$, $z \in \R^{n_z}$, $u \in \mathcal{U}$, $v \in \mathcal{V}$, $e \in \R^{n_y}$, $\hat{y} \in \R^{n_y}$,
	\begin{equation}
		\underline{\alpha}(|x-\psi(z)|) \leq V(x,z) \leq \overline{\alpha}(|\psi^{-R}(x) - z|)	
		\label{eq:ISSassumptionSandwichBound}	
	\end{equation}
	\begin{equation}
		\begin{array}{r}
			\left\langle \nabla V(x,z), (f_p(x,u,v),f_o(z,u,y + e, \hat{y})) \right\rangle  \leq 
			\\[.5em]
			\qquad -\alpha(V(x,z)) + \sum\limits_{i=1}^{N} \gamma_i(|e_i|) + \theta(|v|).
		\end{array}
	\label{eq:ISSassumptionDerivative}
	\end{equation}	
	\hfill $\Box$
	\label{ISSassumption}
\end{ass}
Assumption~\ref{ISSassumption} implies that \eqref{eq:observerAlly} is a global asymptotic observer when $v = 0$ for system \eqref{eq:system} in the sense that \eqref{eq:ISSassumptionSandwichBound} and \eqref{eq:ISSassumptionDerivative} guarantee that, in this case, for any initial condition $x(0) \in \R^{n_x}$, $z(0) \in \R^{n_z}$ and any input $(u,v) \in \mathcal{L_{\mathcal{U}}} \times \{0\}$, 
 the corresponding (maximal) solution $x$ and $z$ to \eqref{eq:system} and \eqref{eq:observerAlly}, if complete\footnote{Completeness of maximal solution will be ensured in Section \ref{Completeness of maximal solutions}}, satisfy $x(t)-\hat{x}(t) \rightarrow 0$ as $t \rightarrow +\infty$, where $\hat{x}(t) = \psi(z(t))$. More precisely, Assumption~\ref{ISSassumption} implies that the estimation error system $x - \hat{x}$ satisfies an input-to-state stability property \cite{sontag2008input} 
with respect to both the network-induced errors $e_i$, which act as additive measurement noises in \eqref{eq:ISSassumptionDerivative}, and to the unknown disturbance input $v$.
In other words, there exist $\beta \in \KL$ and $\gamma \in \Kinf$ such that, for any input $u \in \mathcal{L_U}$ and any disturbance $v \in \mathcal{L_V}$
the corresponding solutions $x$ and $z$ to \eqref{eq:system} and \eqref{eq:observerAlly} respectively, for all $t \geq 0$ satisfy 
 	$|\hat{x}(t) - x(t)| \leq \beta(|\psi^{-R}(x(0)) - z(0)|, t) + \gamma(\sum\limits_{i=1}^{N} \norm{e_i}_{[0,t]} + \norm{v}_{[0,t]}).$
Hence, Assumption~\ref{ISSassumption} is a robustness property of the observer with respect to measurement noises, which is independent of the network.

In view of \cite[Section VI]{astolfi2021stubborn}, the class of observers in \eqref{eq:observerAlly} satisfying Assumption~\ref{ISSassumption} cover various observer designs in the literature, including Luenberger observers for linear systems, various observers for systems with globally Lipschitz vector fields, observers for input affine systems and extended Kalman filters, see \cite{bernard2022observer} and references therein.
%
See \cite{shim2015nonlinear} for further results on input-to-state stability properties for observers.
It is important to notice that for the design of the triggering rule,  that will be presented in Section \ref{TriggeringRuleAndHybridModel}, $\alpha \in \Kinf$ and the Lyapunov function $V$ in Assumption~\ref{ISSassumption} are not needed to be known. Indeed, only $\gamma_i$ is needed and, in addition, we have a lot of freedom regarding the definition of $\gamma_i$, as explained later in Remark~\ref{remarkFreedomISSgains}.
%
Note that we work, for simplicity, with global assumption (see Assumption~\ref{ISSassumption}) but all the analysis could be done in a more local setting (i.e. semi-global, or regional).
\subsection{Problem formulation}
Our goal is to design the local triggering rules to decide when each node $i$ needs to transmit its data to observer~\eqref{eq:observerAlly}, while approximately preserving the properties of observer~\eqref{eq:observerAlly} in the absence of the network  as stated in Assumption~\ref{ISSassumption}.  We assume for this purpose that the $N$ sensors are sufficiently ``smart'' so that they have enough computation capabilities to run a local \textit{scalar} filter, as detailed in the next section. 
%
\section{Design of the triggering rules}\label{TriggeringRuleAndHybridModel}
In the proposed architecture, each sensor node $i \in \{1, \dots, N\}$ has access to its local output measurement $y_i$ and its last transmitted output value $\bar{y}_i$.  
We also introduce a set of local scalar variables $\eta_i \in \R_{\geq0}$, with $i \in \{1, \dots, N\}$. The $\eta_i$-dynamics is, between two successive transmissions of any node and at each transmission of node $i$,
 respectively, given by
\begin{equation}
	\begin{aligned}
		\dot{{\eta_i}} &= -\alpha_i(\eta_i) + c_i\gamma_i(|e_i|) =: \ell_i(\eta_i, e_i)\\
		\eta_i^+ &= b_i \eta_i  \\
		\eta_j^{+} &= \eta_j,  \qquad \ \textnormal{$j \in \{1, \dots, N\}$ with $j \neq i$},
	\end{aligned}
	\label{eq: etaEquation}
\end{equation}
where $\gamma_i \in \Kinf$ comes from Assumption~\ref{ISSassumption}, while $\alpha_i \in \Kinf$, $c_i \geq 0$, $b_i \in [0,1]$ are design functions and parameters. 
%
In particular, equation \eqref{eq: etaEquation} means that when node $i$ transmits, with $i \in \{1,\dots, N\}$, the corresponding $\eta_i$ is updated according to $\eta_i^+ = b_i \eta_i$, while the auxiliary scalar variables $\eta_j$, with $j \in \{1, \dots, N\}$, $j \neq i$, associated to the other sensors are not updated.
\begin{figure}
	\begin{center}
		\tikzstyle{blockB} = [draw, fill=blue!30, rectangle, 
		minimum height=2em, minimum width=3em]  
		\tikzstyle{blockG} = [draw, fill=MyGreen!40, rectangle, 
		minimum height=2em, minimum width=3em]
		\tikzstyle{blockR} = [draw, fill=red!40, rectangle, 
		minimum height=2em, minimum width=3em]
		\tikzstyle{blockO} = [draw,minimum height=1.5em, fill=orange!20, minimum width=4em]
		\tikzstyle{input} = [coordinate]
		\tikzstyle{blockW} = [draw,minimum height=1.5em, fill=white!20, minimum width=2em]
		\tikzstyle{blockWbis} = [draw,minimum height=1.5em, fill=white!20, minimum width=18em]
		\tikzstyle{input1} = [coordinate]
		\tikzstyle{blockCircle} = [draw, circle]
		\tikzstyle{sum} = [draw, circle, minimum size=.3cm]
		\tikzstyle{blockSensor} = [draw, fill=white!20, draw= blue!80, line width= 0.8mm, minimum height=10em, minimum width=13em]
		
		\begin{tikzpicture}[auto, node distance=2cm,>=latex, scale=0.80,transform shape] 
			
			\node [input, name=sensorInput] {};

			\node [blockW, right of=sensorInput, node distance=1.5cm] (sensor) {
				\textbf{Sensor $i$}
			};
			\draw [draw,->] (sensorInput) -- (sensor);
			
			\node [input, right of= sensor, node distance=4.7cm] (sampling) {};
			\node [input, right of= sampling, node distance=0.7cm] (samplingRight) {};
			\node [input, above of= samplingRight, node distance=0.7cm] (samplingFinal) {};
			\draw [draw,-] (sampling) -- (samplingFinal);
			\draw [draw,-] (sensor) -- node  [pos=0.2]{$y_i$}(sampling);
			
			\node [input, above of= sampling, node distance=2.8cm] (ETMitemp) {};
			\node [blockWbis, right of=ETMitemp, node distance=0.3cm] (ETMi) {
				\begin{minipage}{12em} \centering
						\footnotesize Transmit $\bar{y}_i$ when\\[.2em] 
						$\gamma_i (|e_i|) \geq \sigma_i \alpha_i (\eta_i) + \varepsilon_i$\\[.2em]
						where\\[.2em] 
						$\hspace{-3.5em} 
						\left\lbrace 
						\begin{array}{l}
							\dot{{\eta}}_i = -\alpha_i(\eta_i) + c_i\gamma_i (|e_i|) \\
							\eta^{+}_i = b_i\eta_i  \quad \textnormal{when node $i$ transmits}\\
						\end{array}
						\right. $
						
						$\hspace{-4em} \left\lbrace 
						\begin{array}{l}
							\dot{\bar{y}}_i =0, \\ 
							\bar{y}_i^+ = y_i	\quad \ \textnormal{when node $i$ transmits}\\	
						\end{array}
						\right.
						$
				\end{minipage} 
			};
			\node [input, right of= sensor, node distance=1.2cm] (ETMiinput) {};
			\node [input, above of= ETMiinput, node distance=2.8cm] (ETMiinputBis) {};
			\draw [draw,-] (ETMiinput) -- (ETMiinputBis);
			\draw [draw,->] (ETMiinputBis) --  (ETMi);
			\node [input, below of= ETMi, node distance=2.4cm] (ETMiarrow) {};
			\draw [draw,->] (ETMi) --  (ETMiarrow);
			\node [input, above of= ETMi, node distance=1.65cm, right = 0.72cm] (ETMiName) {};
			\draw [] (ETMiName) -- node  [pos=0.5]{\textbf{ETM \textit{i}}} 	(ETMiName);

			\node [input, right of= samplingRight, node distance=3.2cm] (finalPoint) {};
			\draw [draw,->] (samplingRight) -- node  [pos=0.7]{$\bar y_i$}  (finalPoint);
		\end{tikzpicture}
	\end{center}
	\caption{Event triggering mechanism (ETM) of node $i$, $i \in \{1, \dots, N\}$}
	\label{Fig:sensorNode}
\end{figure}
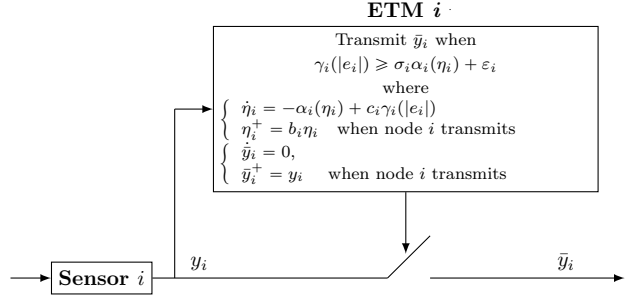
The auxiliary scalar variable $\eta_i$ is used to define the triggering instants for sensor node $i$. Indeed, sensor $i$, with $i \in \{1, \dots, N\}$, transmits its output measurement only when the condition
\begin{equation}
	\gamma_i(|e_i|) \geq \sigma_i\alpha_i(\eta_i) + \varepsilon_i
	\label{eq:triggeringRule}
\end{equation} 
is satisfied, where $\sigma_i \geq 0$ and $\varepsilon_i > 0$ are additional design parameters, as summarized in Figure~\ref{Fig:sensorNode}. Note that the parameter $\varepsilon_i$ is essential to avoid the Zeno phenomena. Indeed, we will show in Section~\ref{IET},  under mild extra conditions, that there exists a strictly positive minimum time between any two transmissions of the same sensor node, which vanishes when $\varepsilon_i =0$. 
\begin{rem}\label{remarkFreedomISSgains}
	To design the triggering mechanism it is not necessary to know $\alpha \in \Kinf$ and the Lyapunov function $V$ in Assumption~\ref{ISSassumption} in view of \eqref{eq: etaEquation}-\eqref{eq:triggeringRule}: only $\gamma_i$ is needed, and, as a result, there is a lot of freedom regarding the definition of $\gamma_i$.
	Indeed, if Assumption~\ref{ISSassumption} is satisfied with some $\gamma_1, \dots, \gamma_N \in \Kinf$, then Assumption~\ref{ISSassumption} holds with any $\tilde{\gamma}_1, \dots, \tilde{\gamma}_N \in \Kinf$ verifying $\gamma_i(r) = O(\tilde{\gamma}_i(r))$ as $r \rightarrow +\infty$ with a different $V$ and a different $\alpha$ in view of Lemma~\ref{LemmaMultipleChangeOfSupplyRate} in the appendix. This implies for instance that, when Assumption~\ref{ISSassumption} holds with $\gamma_i$ quadratic for all $i \in \{1, \dots, N\}$, the $\gamma_i$'s can be replaced by \textit{any} quadratic function in \eqref{eq: etaEquation}-\eqref{eq:triggeringRule}. We will exploit this property in the example in Section~\ref{Example}. \hfill $\Box$
\end{rem}

We write $\eta:=(\eta_1,\dots,\eta_N) \in \R^N$ and we define the overall state as $q:= (x,z,e,\eta) \in \mathcal{Q} := \R^{n_x} \times \R^{n_z} \times \R^{n_y} \times \R_{\geq0}^N$ and the overall input $w:=(u,v) \in \mathcal{W} := \mathcal{U} \times \mathcal{V}$. 
We obtain the hybrid model
\begin{equation}
	\left\lbrace 
	\begin{aligned}
		\dot{q} &= F(q,w), \ \ \ \ \ &&q \in \mathcal{C}\\
		q^{+} &\in G(q), \ \ \ \ \ &&q \in \mathcal{D}.
	\end{aligned}
	\right.
	\label{eq:HybridSystem}
\end{equation}
where the flow map $F$ is defined as, for any $q \in \mathcal{C}$ and any $w 	\in \mathcal{W}$, $F(q,w) :=  \big(f_p(x,w), f_o(z, u, h(x), h(\psi(z))), g (x,w), \ell(\eta, e)\big),$ where $ \ g(x,w) := $ $ (g_{1}(x,w), \dots, g_N(x,w))$ with $g_i$ in  \eqref{eq:samplingErrorContinuous} 
and $\ell(\eta, e) :=$ $(\ell_1(\eta_1,e_1), \dots, \ell_N(\eta_N,e_N))$ with $\ell_i$ in \eqref{eq: etaEquation}.
The flow set $\mathcal{C}$ is defined as 
\begin{equation}
	\mathcal{C} := \bigcap_{i=1}^{N} \mathcal{C}_i
	\label{eq:FlowSet}
\end{equation}
with 
\begin{equation}
	\mathcal{C}_i := \left\{q \in \mathcal{Q}: \gamma_i(|e_i|) \leq \sigma_i \alpha_i( \eta_i) + \varepsilon_i \right\},
	\label{eq:FlowSet_i}
\end{equation}
for any $i \in \{1, \dots, N\}$. On the other hand, the jump set $\mathcal{D}$ is defined as
\begin{equation}
	\mathcal{D} := \bigcup_{i=1}^{N} \mathcal{D}_i
	\label{eq:JumpSet}
\end{equation}
with 
\begin{equation}
	\mathcal{D}_i := \left\{q \in \mathcal{Q}: \gamma_i(|e_i|) \geq \sigma_i \alpha_i( \eta_i) + \varepsilon_i \right\},
	\label{eq:JumpSet_i}
\end{equation}
for any $i \in \{1, \dots, N\}$.
Sets $\mathcal{C}$ and $\mathcal{D}$ in \eqref{eq:FlowSet}-\eqref{eq:JumpSet_i} are such that a transmission is triggered whenever one of the conditions $ \gamma_i(|e_i|) \geq \sigma_i \alpha_i( \eta_i) + \varepsilon_i$ is satisfied by at least one sensor node, as illustrated in Figure~\ref{Fig:sensorNode}. These conditions may be verified simultaneously by different sensor nodes. In this case, several jumps may occur immediately one after the other, with no flow in between. 
It is important to note that we recover the 
absolute threshold triggering rule, as in, e.g., \cite{etienne2017periodic, etienne2016event, etienne2017asynchronous, tong2020finite}, as a special case, by taking $\sigma_i = 0$ for all $i \in \{1, \dots, N\}$. 

The set-valued jump map $G$ in \eqref{eq:HybridSystem} is defined as, for any $q \in \mathcal{D}$,
\begin{equation}
	G(q) := \bigcup_{i = 1}^{N}G_{i}(q),
	\label{eq:jumpMap}
\end{equation}
with 
\begin{equation}
	G_{i}(q) := 	\left\lbrace 
	\begin{aligned}
		&\left( 
		\begin{array}{c}
			x\\
			z\\
			\Lambda_i e \\
			(b_i(I_N - \Gamma_i) + \Gamma_i)\eta
		\end{array}
		\right) 
		\ \ \ \ &&q \in \mathcal{D}_i\\
		& \qquad \qquad \quad \emptyset &&q \notin \mathcal{D}_i,
	\end{aligned}
	\right.
	\label{eq:jumpMap_i_Esplicit}
\end{equation}
where $\Lambda_i$ is the block diagonal matrix of dimension $n_y$ with $N$ blocks, where the $i$-th block is $0_{n_{y_i} \times n_{y_i}}$, while all the other blocks are $I_{n_{y_j}}$, for all $i \in \{1, \dots, N\}$, $j \in \{1, \dots, N\}$, with $j \neq i$.  Moreover, 
$\Gamma_i$ is the diagonal matrix of dimension $N$ with all elements on the diagonal being equal to $1$ except for the $i$-th element, which is $0$, for $i \in \{1, \dots, N\}$.  
The set $\mathcal{D}_i$ corresponds to the region of the state space where a triggering of node $i$ is allowed.
Indeed, a jump in \eqref{eq:HybridSystem} corresponds to a transmission of one current output $y_i$ to the observer. In this case $x^+ = x$, $z^+ = z$, $e_i^+ = 0$,  $e_j^+ = e_j$, $\eta_i^+ = b_i\eta_i$ and $\eta_j^+ = \eta_j$ for $j \in \{1, \dots, N\}$ with $j \neq i$. The empty set in \eqref{eq:jumpMap_i_Esplicit} essentially means that we consider the jump map $G_i$ only when its argument is in the jump set $\mathcal{D}_i$. Indeed, in our setting, each sensor performs its output transmission, according to $G_i$, independently of the other sensors and the transmission does not affect the other sensor nodes.  However, this notation is useful because we also have to define $G_i(q)$ when $q \notin \mathcal{D}_i$ in view of the definition of the jump set $\mathcal{D}$ in \eqref{eq:JumpSet}-\eqref{eq:JumpSet_i}. 

We are ready to proceed with the design of $\alpha_i$, $\sigma_i$, $c_i$, $\varepsilon_i$, $b_i$ in \eqref{eq: etaEquation}-\eqref{eq:triggeringRule} and the stability analysis of system~\eqref{eq:HybridSystem}.
\section{Stability guarantees}\label{StabilityGuarantees}
We first present Lyapunov properties in Sections~\ref{LyapunovPropertiesSubsection}, then we derive stability guarantees in Section~\ref{StabilitySolutionsSubsection}. The corresponding proofs are provided in Section~\ref{StabilityProofsSubsection}.
\subsection{Lyapunov properties}\label{LyapunovPropertiesSubsection}
The objective of this section is to prove that the proposed event-triggered observer satisfies a uniform global practical stability property. In particular, for any $\nu> 0$ we can tune the triggering mechanism such that 
	there exist $\beta^\star \in \KL$ and $\gamma^\star \in \Kinf$ such that, for any input $w \in \mathcal{L}_\mathcal{W}$, any corresponding solution $q$ to \eqref{eq:HybridSystem}-\eqref{eq:jumpMap_i_Esplicit}, for all $(t,j) \in \dom q$, 
	satisfies
	\begin{equation}
		\begin{aligned}
			|x(t,j) -\hat{x}(t,j)| \leq & \beta^\star(|(\psi^{-R}(x(0,0)) -z(0,0), \eta(0,0))|,t)  \\ 
			&+ \gamma^\star(\nu +\norm{v}_{[0,t]}).
		\end{aligned}
		\label{eq:ISS_KL}
	\end{equation}
For this purpose, we first prove a Lyapunov stability property for the overall system \eqref{eq:HybridSystem} in the next theorem.	
%

\begin{thm}
	Suppose Assumptions~\ref{assumption 1}-\ref{ISSassumption} hold and consider the hybrid model \eqref{eq:HybridSystem}-\eqref{eq:jumpMap_i_Esplicit}. For any $\nu > 0$,  select $\sigma_i^\star > 0$, $c_i^\star \geq 0$ such that $\sigma_i^\star c_i^\star <1$ and  $d_i > d_i^\star$ where  $\displaystyle d_i^\star:= \frac{\sigma_i^\star}{1-\sigma_i^\star c_i^\star} >0$ and select $\varepsilon_i^\star >0$ such that $\displaystyle \sum\limits_{i = 1}^{N}(1 + d_ic_i^\star)\varepsilon_i^\star \leq \nu$, 
	for all $i\in \{1, \dots, N\}$. 
	Define 
	\begin{equation}
		U(q):=V(x,z) + \sum\limits_{i = 1}^{N} d_i \eta_i,
		\label{eq:LyapunovDefinitionU}
	\end{equation} 
	for any $q \in \mathcal{Q}$. Then, there exist $\underline{\alpha}_U$, $\overline{\alpha}_U$ $\in \Kinf$ such that for any $\alpha_i \in \Kinf$ in \eqref{eq: etaEquation}, $\sigma_i \in [0, \sigma_i^\star]$, $c_i \in [0, c_i^\star]$, $\varepsilon_i \in (0, \varepsilon_i^\star]$ and  $b_i \in [0,1]$, for all $i \in \{1, \dots, N\}$,  the following properties hold. 
	\begin{enumerate}[label=(\roman*),leftmargin=.7cm]
		\item 
		For any $q \in \mathcal{Q}$, 
		\begin{equation}
			\underline{\alpha}_U(|(x-\psi(z), \eta)|) \leq U(q) \leq \overline{\alpha}_U(|(\psi^{-R}(x)-z, \eta)|).
			\label{eq:sandwichBoundU}
		\end{equation}
		\item For any $q \in \mathcal{C}$ and any $w \in \mathcal{W}$, 
		\begin{equation}
			\begin{array}{l}
			\left\langle \nabla U(q), F(q,w) \right\rangle  \leq \\[.5em] \qquad -\alpha(V(x,z)) - \sum\limits_{i=1}^{N} \delta_i \alpha_i(\eta_i) + \nu + \theta(|v|),
						\end{array}
			\label{eq:stabilityTheoremFlow}
		\end{equation}
		where $\alpha, \theta \in \Kinf$ come from Assumption~\ref{ISSassumption}, and $\delta_i := d_i -\sigma_i^\star(1 + d_ic_i^\star) >0$. 
		\item For any $q \in \mathcal{D}$, for any $\mathfrak{g} \in G(q)$,
		\begin{equation}
			U(\mathfrak{g}) \leq U(q).
			\label{eq:stabilityTheoremJump}
		\end{equation}
	\end{enumerate}\hfill $\Box$	
	\label{theoremLyapunov}
\end{thm}
Theorem~\ref{theoremLyapunov} shows the existence of a Lyapunov function $U$ for system \eqref{eq:HybridSystem}-\eqref{eq:jumpMap_i_Esplicit}, which guarantees a uniform practical stability property, where the adjustable parameter is $\nu$. 
 %
The conditions of Theorem~\ref{theoremLyapunov} can always be ensured. Indeed, we just need to select $\sigma_i^\star$ and $c_i^\star$ such that $\sigma_i^\star c_i^\star <1$, for all $i \in \{1, \dots, N\}$, which is always possible and then all the other parameters can be selected such that conditions in Theorem~\ref{theoremLyapunov} hold. 
Moreover, $\nu$ in \eqref{eq:stabilityTheoremFlow} can be taken arbitrary small. However, typically the smaller $\nu$ is selected, the higher the number of transmissions required. 
In Theorem~\ref{theoremLyapunov}, we first fix $\nu$ and then we present how to select the design parameters in order to obtain the Lyapunov properties in \eqref{eq:sandwichBoundU}-\eqref{eq:stabilityTheoremJump}. 
An alternative approach is to select $\sigma_i$ and $c_i$ such that $\sigma_ic_i < 1$ for all $i \in \{1, \dots, N\}$, and then, by simply selecting $b_i \in [0,1]$, and any positive value for $\varepsilon_i$, any $\alpha_i \in \Kinf$, for all $i \in \{1, \dots, N\}$, \eqref{eq:sandwichBoundU}-\eqref{eq:stabilityTheoremJump} hold for some strictly positive $\nu$. The selection of the design parameters in the example in Section~\ref{Example} is done exploiting this second strategy. 
\begin{rem} \label{decayRateRemark}
	In absence of network, Lyapunov function $V$ decays with a rate $\alpha \in \Kinf$ along the solutions to \eqref{eq:system} and \eqref{eq:observerAlly} according to Assumption~\ref{ISSassumption}. We can ensure any decay rate $\alpha_U \in \Kinf$ such that $\alpha_U \leq \alpha$ on flows for the Lyapunov function $U$ along the solutions to \eqref{eq:HybridSystem}-\eqref{eq:jumpMap_i_Esplicit} on any given compact set by suitably selecting $\alpha_i$ in \eqref{eq: etaEquation}, for all $i \in \{1, \dots, N\}$. 
	The result is global in some special cases, like when $\alpha \in \Kinf$ is subadditive, i.e. $\alpha(s_1) + \alpha(s_2) \geq \alpha(s_1 + s_2)$, for all $s_1, s_2 \geq 0$, or when $\alpha \in \Kinf$ is uniformly continuous, see also Theorem~\ref{theoremLyapunovSolutionLinear} and Lemma~\ref{decayRateLemma} in the Appendix. 
	\hfill $\Box$ 
\end{rem}

\subsection{Uniform global practical stability property}\label{StabilitySolutionsSubsection}
Based on Theorem~\ref{theoremLyapunov}, we prove that the event-triggered observer satisfies a global practical stability property.

\begin{prop}
	Consider system \eqref{eq:HybridSystem}-\eqref{eq:jumpMap_i_Esplicit} 
	and suppose Assumptions~\ref{assumption 1}-\ref{ISSassumption} hold. For any $\nu > 0$, select $\alpha_i$, $\sigma_i$, $c_i$, $\varepsilon_i$, $d_i$ and $b_i$ as in Theorem~\ref{theoremLyapunov} for all $i\in \{1, \dots, N\}$. Then there exist $\beta_U \in \KL$ and $\gamma_U$, $ \theta_U \in \Kinf$, all independent of $\nu$, such that, for any input $w \in \mathcal{L_W}$, any solution  $q$ 
	satisfies for all $(t,j) \in \dom q$,
	\begin{equation}
		\begin{array}{l}
		V(x(t,j), z(t,j)) + \sum\limits_{i = 1}^{N} d_i\eta_i(t,j) \\[0.5em]
		\qquad \leq \beta_U(V(x(0,0), z(0,0)) + \sum\limits_{i = 1}^{N} d_i\eta_i(0,0), t) \\[0.5em]
		\qquad \quad + \gamma_U(\nu + \theta(\norm{v}_{[0,t]})).
		\end{array}
		\label{eq:LyapunovSolutionDecreasing}
	\end{equation}
	with $\theta \in \Kinf$ from Assumption~\ref{ISSassumption}.
	\hfill $\Box$
	\label{TheoremLyapunovSolution}
\end{prop}

From \eqref{eq:sandwichBoundU} and Proposition~\ref{TheoremLyapunovSolution}, we have that the estimation error $x - \hat{x}$ satisfies a uniform global practical stability property in the sense that \eqref{eq:ISS_KL} holds.
Moreover, \eqref{eq:LyapunovSolutionDecreasing} also ensures that the $\eta_i$ components, with $i \in \{1, \dots, N\}$, are bounded and converge to a neighborhood of the origin. 

\begin{rem}\label{RemarkWhyOnlyPractical} 
	To ensure an asymptotic stability property for the estimation error system, as opposed to a practical one as in Proposition~\ref{TheoremLyapunovSolution}, we argue that a different set-up would be needed, which would require to implement a copy of the observer at each node. Indeed, a typical way to ensure an asymptotic stability property for the estimation error system when emulating an observer of the form of \eqref{eq:observerAlly} is not to only hold the plant output $y$ as we do in \eqref{eq:observer} but the output estimation error $\bar y - y$ see e.g., \cite{ferrante2018mathcal, postoyan2011framework, postoyan2014tracking}. In this case, the network-induced error associated to node $i$ becomes $(\bar y_i - \hat{\bar{ y}}_i) - (y_i-\hat y_i)$. 
	 Hence, for the local triggering rule $i$ to evaluate this network-induced error, it would need to know  $\hat{y}_i$, which can only be done by implementing a local copy of the observer at node $i$ to generate  $\hat{y}_i$. 
	Because our goal is precisely not to rely on a copy of the observer at each node, as explained in the introduction, the triggering rules we present do not rely on $\hat y_i$, but only on $y_i$ (and $\eta_i$), which leads to a practical stability property.
	\hfill $\Box$
\end{rem}

As mentioned before, we do not need to know $\alpha \in \Kinf$ and $V$ to design the triggering conditions such that the results in Theorem~\ref{theoremLyapunov} and in Proposition~\ref{TheoremLyapunovSolution} hold. However, the knowledge of $\alpha \in \Kinf$ is useful when we want to recover the decay rate $\alpha \in \Kinf$ of the Lyapunov function along solutions in absence of network, as formalized below for the case where Assumption~\ref{ISSassumption} holds with $\alpha$ linear, see also Remark~\ref{decayRateRemark}. 

\begin{thm}	
	Consider system \eqref{eq:HybridSystem}-\eqref{eq:jumpMap_i_Esplicit} 
	and suppose Assumption~\ref{assumption 1} holds and Assumption~\ref{ISSassumption} is satisfied with $\alpha(s) = as$, for any $s \geq 0$ with $a>0$. For any $a_U \in (0, a]$ and $\mu >0$ select $\alpha_i$, $c_i$, $\sigma_i$, $\varepsilon_i$ and $b_i$ as follows for all $i \in \{1, \dots, N\}$.
	\begin{enumerate}[label=(\roman*),leftmargin=.7cm]
		\item $c_i \in [0, c_i^{\star}]$ and $\sigma_i \in [0, \sigma_i^{\star}] $, where $c_i^{\star} \geq 0$ and $\sigma_i^{\star} > 0$ are such that $\sigma_i^{\star}c_i^{\star} < 1$, for all $i \in \{1, \dots, N\}$.
		\item $\alpha_i(s) = a_is$ for any $s \geq 0$ 
		with $a_i \geq a_i^{\star}$ and $a_i^{\star}> 0$ such that $ a_i^{\star}>\frac{{a}_U}{1-\sigma_i^{\star}c_i^{\star}}$, for all $i \in \{1, \dots, N\}$.
		\item $b_i \in [0,1]$,  for all $i \in \{1, \dots, N\}$.
		\item  $\varepsilon_i \in (0, \varepsilon^\star_i]$ for all $i \in \{1, \dots, N\}$ and $\displaystyle \varepsilon_1^\star + \ldots + \varepsilon_N^\star \leq \frac{{a}_U\mu}{1 + \varsigma}$ with $\varsigma:= \max\{d_1c_1^\star, \dots, d_Nc_N^\star\}$, where $ d_i:= \sigma_i^{\star} \big (1-\sigma_i^{\star} c_i^{\star} -\frac{a_U}{a_i^{\star}} \big )^{-1} > 0$, for all $i \in \{1, \dots, N\}$.
	\end{enumerate}	
	Then, for $U$ defined in \eqref{eq:LyapunovDefinitionU} with $d_i$ selected as in item \textit{(iv)}, which satisfies the condition stated in Theorem~\ref{theoremLyapunov}, for all $i \in \{1, \dots, N\}$, for any solution $q$ with input $w \in \mathcal{L_W}$ and any $(t,j) \in \dom q$,  
		$	V(x(t,j), z(t,j)) + \sum\limits_{i = 1}^{N} d_i \eta_i(t,j) 
			\leq e^{-a_Ut}	(V(x(0,0), z(0,0)) + \sum\limits_{i = 1}^{N} d_i \eta_i(0,0)) 
			+ \mu + \frac{1}{a_U}\theta(\norm{v}_{[0,t]})$.
	\label{theoremLyapunovSolutionLinear}\hfill $\Box$
\end{thm}
	Theorem~\ref{theoremLyapunovSolutionLinear} guarantees that it is always possible to recover the same decay rate of the Lyapunov function along solutions in absence of network when the observer satisfies Assumption~\ref{ISSassumption} with $\alpha$ linear. 
	In particular, with Theorem~\ref{theoremLyapunovSolutionLinear}
	we guarantee, in presence of network,  a convergence rate $a_U \in (0, a]$ for $U(q) = V(x,z) +  \sum\limits_{i = 1}^{N} d_i \eta_i$ along solutions to \eqref{eq:HybridSystem}-\eqref{eq:jumpMap_i_Esplicit}, which can therefore be equal to the decay rate $a$ of $V$ in absence of network. 
	
	It is important to notice that many observers in the literature satisfy Assumption~\ref{ISSassumption} with a linear $\alpha$, see \cite{astolfi2021stubborn}.
	Moreover, it is always possible to ensure the conditions in Theorem~\ref{theoremLyapunovSolutionLinear}, like in Theorem~\ref{theoremLyapunov}. Indeed, selecting $\sigma_i^\star$ and $c_i^\star$ such that $\sigma_i^\star c_i^\star < 1$ for all $i \in \{1, \dots, N\}$, which is always possible, we have that all the other parameters can be always chosen such that items \textit{(ii)-(iv)} of Theorem~\ref{theoremLyapunovSolutionLinear} are satisfied. 

\subsection{Proofs}\label{StabilityProofsSubsection}

\subsubsection{Proof of Theorem~\ref{theoremLyapunov}}
Let all conditions of Theorem~\ref{theoremLyapunov} hold. 

Item \textit{(i)} of Theorem~\ref{theoremLyapunov} follows from \eqref{eq:ISSassumptionSandwichBound} and \eqref{eq:LyapunovDefinitionU} and the application of \cite[Lemma 4]{wang2019periodic}. In particular, it holds with $ \underline{\alpha}_U(s):= \min\left\{\underline{\alpha}\left(\frac{s}{N+1}\right), d_1 \frac{s}{N+1}, \dots, \right.$ $\left.  d_N \frac{s}{N+1}\right\}$ and  $ \overline{\alpha}_U(s) = \overline{\alpha}(s)  + \sum\limits_{i = 1}^{N} d_i s$ for any $s \geq 0$. 
We now prove that item \textit{(ii)} of Theorem~\ref{theoremLyapunov} holds. 
Let $\displaystyle q \in \mathcal{C}$ and $w \in \mathcal{W}$. In view of \eqref{eq:ISSassumptionDerivative}, \eqref{eq: etaEquation} and \eqref{eq:LyapunovDefinitionU}, 
		$ \left\langle \nabla U(q), F(q,w) \right\rangle  \leq -\alpha (V(x,z)) +\sum\limits_{i = 1}^{N}\gamma_i(|e_i|) + \theta(|v|)  +\sum\limits_{i = 1}^{N}d_i(-\alpha_i(\eta_i) + c_i \gamma_i(|e_i|)) = -\alpha (V(x,z)) - \sum\limits_{i = 1}^{N}d_i\alpha_i(\eta_i) 
         +\sum\limits_{i = 1}^{N}(1 + d_ic_i)\gamma_i(|e_i|) + \theta(|v|).
$
Since $\displaystyle q \in \mathcal{C}$, we have from \eqref{eq:FlowSet_i} that $\gamma_i(|e_i|) \leq \sigma_i \alpha_i(\eta_i) + \varepsilon_i$ for all $i \in \{1, \dots, N\}$. Hence, the next inequality holds
$
		\left\langle \nabla U(q), F(q,w) \right\rangle  
		 \leq -\alpha (V(x,z)) - \sum\limits_{i = 1}^{N}d_i\alpha_i(\eta_i) 
		+ \sum\limits_{i = 1}^{N}(1 + d_ic_i)(\sigma_i \alpha_i(\eta_i) + \varepsilon_i) + \theta(|v|) 	
		 = -\alpha (V(x,z))- \sum\limits_{i = 1}^{N}(d_i -\sigma_i(1 + d_ic_i))\alpha_i(\eta_i) 
	    + \sum\limits_{i = 1}^{N}(1 + d_ic_i)\varepsilon_i + \theta(|v|) .
$
Due to the conditions $\sigma_i \in [0, \sigma_i^\star], c_i \in [0,c_i^\star]$ and $\varepsilon_i \in (0, \varepsilon_i^\star]$ in Theorem~\ref{theoremLyapunov}, 
$
		\left\langle \nabla U(q), F(q,w) \right\rangle  
		\leq -\alpha (V(x,z))- \sum\limits_{i = 1}^{N}(d_i -\sigma_i^\star(1 + d_ic_i^\star))\alpha_i(\eta_i)   
		+ \sum\limits_{i = 1}^{N}(1 + d_ic_i^\star)\varepsilon_i^\star + \theta(|v|).
$
Using the definitions of $\delta_i$ in item $(ii)$ of Theorem~\ref{theoremLyapunov} and the fact that $ \displaystyle \nu \geq \sum\limits_{i = 1}^{N}(1 + d_ic_i^\star)\varepsilon_i^\star $, we obtain 
$
	\left\langle \nabla U(q), F(q,w) \right\rangle  
	\leq -\alpha (V(x,z))- \sum\limits_{i = 1}^{N}\delta_i\alpha_i(\eta_i) + \nu + \theta(|v|),
$
where $\delta_i$ is strictly positive for any $i \in \{1, \dots, N\}$ as $d_i > d_i^\star$ and $\sigma_i^\star c_i^\star < 1$. The proof of item \textit{(ii)} is complete.

We finally prove that item \textit{(iii)} of Theorem~\ref{theoremLyapunov} is satisfied.
 Let $\displaystyle q \in \mathcal{D}$, in view of \eqref{eq: etaEquation} and \eqref{eq:jumpMap_i_Esplicit} and since $b_i \in [0,1]$ for all $i \in \{1, \dots, N\}$, for any $\mathfrak{g} \in G(q)$, there exists $k\in \{1, \dots, N\}$ such that $\mathfrak{g} \in G_k(q)$, hence
$ U(\mathfrak{g}) = V(x,z) + \sum\limits_{\substack{i = 1 \\ i \neq k}}^{N}d_i\eta_i + d_kb_k\eta_k 
\leq V(x,z) + \sum\limits_{i = 1}^{N}d_i\eta_i
= U(q),$
which concludes the proof of item \textit{(iii)}. 
\subsubsection{Proof of Proposition~\ref{TheoremLyapunovSolution}}
Consider the Lyapunov function $U$ defined in \eqref{eq:LyapunovDefinitionU}. 
From item  \textit{(ii)} of Theorem~\ref{theoremLyapunov} and \cite[Lemma 4]{wang2019periodic}, we derive that for any $q \in \mathcal{C}$ and $w\in \mathcal{W}$, 
$
		\left\langle \nabla U(q), F(q,w) \right\rangle
		 \leq 
		 -\alpha_U(U(q)) + \nu +\theta(|v|), 
$
where $ \alpha_U(s):= \min\left\{ \alpha\left(\frac{s}{2}\right), {\alpha}_{\eta} \left(\frac{s}{2}\right)\right\}$ and $\alpha_{\eta}(s):= \min\left\{\delta_1\alpha_1\left(\frac{s}{\bar{d}N}\right), \right.$ $\left. \dots,   \delta_N\alpha_N\left(\frac{s}{\bar{d}N}\right) \right\}$, with $\bar{d}:= \max\{d_1, \dots, d_N\}$.
%
%
Hence, given $\zeta \in (0,1)$, when $\nu + \theta(|v|) \leq (1-\zeta) \alpha_U(U(q))$, 
\begin{equation}
	\begin{array}{l}
		\left\langle \nabla U(q), F(q,w) \right\rangle 
		\leq -\zeta\alpha_U(U(q)). 
	\end{array}
	\label{eq:LyapunovUFlowsVectors2}
\end{equation}
We then follow similar steps as in  \cite[proof of Theorem 3.18]{goedel2012hybrid1}. Let $w \in \mathcal{L_{\mathcal{W}}}$ and $q$ be a solution to system \eqref{eq:HybridSystem}-\eqref{eq:jumpMap_i_Esplicit}. 
Pick any $(t,j) \in \dom q$ and let $0 = t_0 \leq t_1 \leq \dots \leq t_{j+1} = t$ satisfy $\dom  q \cap ([0,t]\times \{0,1,\dots,j\}) = \bigcup_{k=0}^j [t_k, t_{k+1}] \times \{k\}$. For each $k \in \{0,\dots,j\}$ and almost all $s \in [t_k, t_{k+1}]$, $q(s,k) \in \mathcal{C}$. In view of \eqref{eq:LyapunovUFlowsVectors2}, applying \cite[pages 19-21]{isidori1999nonlinear}, 
there exists $\beta_U \in \KL, \gamma_U \in \Kinf$ such that 
$
		U(q(s,k)) 
 \leq \beta_U(U(q(t_k,k)), s-t_k) +  \gamma_U(\nu + \theta(\norm{v}_{[t_k, s]}))
$
for all $s \in [t_k, t_{k+1}]$, for all $k \in \{0,\dots,j\}$.
Consequently, we have, for any $k \in \{0,\dots,j\}$,
\begin{equation}
	\begin{aligned}
	U(q(t_{k+1},k)) &\leq \beta_U((U(q(t_k,k)), t_{k+1}-t_k) \\ 
	&+ \gamma_U(\nu + \theta(\norm{v}_{[0,t_{k+1}]}))  
	\end{aligned}
\label{eq:equationProofCorollaryKLbound6}
\end{equation}
On the other hand, from item \textit{(iii)} of Theorem~\ref{theoremLyapunov}, for each $k\in \{1,\dots,j\}$, 
\begin{equation}
	U(q(t_k,k)) - U(q(t_k,k-1)) \leq 0 \ \ \ \forall k \in \{1,\dots,j\}.
	\label{eq: solutionsJumps}
\end{equation}
From \eqref{eq:equationProofCorollaryKLbound6} and \eqref{eq: solutionsJumps}, we deduce that for any $(t,j) \in \dom q$,
\begin{equation}
	U(q(t,j)) \leq \beta_U(U(q(0,0)),t) + \gamma_U(\nu + \theta(\norm{v}_{[0,t]})).
\end{equation}
Using the $U$ definition in \eqref{eq:LyapunovDefinitionU}, we obtain \eqref{eq:LyapunovSolutionDecreasing}, which concludes the proof. 

\subsubsection{Proof of Theorem~\ref{theoremLyapunovSolutionLinear}}
Let all conditions of Theorem~\ref{theoremLyapunovSolutionLinear} hold and consider the Lyapunov function $U$ defined in \eqref{eq:LyapunovDefinitionU} with $d_i$ satisfying item (\textit{iv}) of Theorem~\ref{theoremLyapunovSolutionLinear}. Note that $d_i$ satisfies the condition $d_i > d_i^\star$ in Theorem~\ref{theoremLyapunov}.
As $\alpha (s) = as$ and $\alpha_i(s) = a_is$ for any $s \geq 0$, for all $i \in \{1, \dots N\}$,
by following the steps of the proof of Theorem~\ref{theoremLyapunov}, we derive that 
 for any $q \in \mathcal{C}$ and $w \in \mathcal{W}$,
$
		\left\langle \nabla U(q), F(q,w) \right\rangle \leq 
		-a V(x,z) - \sum\limits_{i=1}^{N} \delta_i a_i \eta_i + \sum\limits_{i=1}^{N}(1 + d_ic_i^\star)\varepsilon_i^\star + \theta(|v|).
$
Defining $ a_{\eta}:= \min\left\{\frac{\delta_1a_1}{d_1}, \dots, \frac{\delta_Na_N}{d_N}\right\} >0$, we obtain
\begin{equation}
	\begin{array}{l}
		\left\langle \nabla U(q), F(q,w) \right\rangle \\[.5em]
		\qquad \leq -a V(x,z) - a_{\eta}\sum\limits_{i=1}^{N} d_i\eta_i + \sum\limits_{i=1}^{N}(1 + d_ic_i^\star)\varepsilon_i^\star+ \theta(|v|)\\[.5em]
		\qquad \leq -\min\{a, a_{\eta}\}(V(x,z) + \sum\limits_{i=1}^{N} d_i\eta_i)  \\[.5em] 
		\qquad \quad + \sum\limits_{i=1}^{N}(1 + d_ic_i^\star)\varepsilon_i^\star + \theta(|v|)\\[.5em]
		\qquad = -\min\{a, a_{\eta}\}U(q)  + \sum\limits_{i=1}^{N}(1 + d_ic_i^\star)\varepsilon_i^\star+ \theta(|v|)\\[.5em]
		\qquad \leq -a_U U(q) + \sum\limits_{i=1}^{N}(1 + d_ic_i^\star)\varepsilon_i^\star+ \theta(|v|),
	\end{array}
	\label{eq:MinAlphaULinear}
\end{equation}
where the last inequality comes from the choice of parameters. Indeed, when $\min\{a, a_{\eta}\} = a$, then $-\min\{a, a_{\eta}\} = -a \leq -a_U$. Conversely, when $\min\{a, a_{\eta}\}$ $ = a_{\eta} =  \min\left\{\frac{\delta_1a_1}{d_1}, \dots, \frac{\delta_Na_N}{d_N}\right\}$, we have from the definition of $\delta_i$ in item (\textit{ii}) of Theorem~\ref{theoremLyapunov}, for all $i \in \{1, \dots, n\}$, $
		-\frac{\delta_i a_i}{d_i} = - (d_i -\sigma_i^\star(1+d_ic_i^\star))\frac{a_i}{d_i}
		 \leq  - (d_i -\sigma_i^\star(1+d_ic_i^\star))\frac{a_i^\star}{d_i}
		= -\big(1 - \sigma_i^\star\big(\frac{1}{d_i} + c_i^\star\big)\big) a_i^\star$
and since  $ d_i = \sigma_i^\star\big(1 - \sigma_i^\star c_i^\star - \frac{a_U}{a_i^\star}\big)^{-1}$, we derive that $ -\frac{\delta_i a_i}{d_i} \leq -a_U$.
Therefore \eqref{eq:MinAlphaULinear} holds and since $ \sum\limits_{i=1}^{N}\varepsilon_i^{\star} \leq \frac{a_U\mu}{1 + \varsigma}$,  with $\varsigma= \max\{d_1c_1^\star, \dots, d_Nc_N^\star\}$, we have
\begin{equation}
	\begin{array}{l}
		\left\langle \nabla U(q), F(q,w) \right\rangle 
		\leq -a_U U(q) + (1 + \varsigma)\sum\limits_{i=1}^{N}\varepsilon_i^\star + \theta(|v|)\\[.5em]
		\qquad \qquad \qquad \quad \; \; \, \leq -a_U U(q) + a_U\mu  + \theta(|v|).
	\end{array}
	\label{eq:LyapunovUFlowsVectorsLinear}
\end{equation}
The desired result is obtained by following similar lines as in the proof of Proposition~\ref{TheoremLyapunovSolution}.
\section{Properties of the solution domains} \label{SolutionDomainSection}
We present in this section the properties of the domain of the solutions to system~\eqref{eq:HybridSystem}-\eqref{eq:jumpMap_i_Esplicit}. In Section~\ref{Completeness of maximal solutions}, we show that maximal solutions are complete, while in Section~\ref{IET} we prove that the time between any two consecutive transmissions of each sensor node is lower-bounded by a uniform strictly positive constant. Finally, we show in Section~\ref{stopTransmissionSubsection} that  
the triggering condition associated to node $i$ stops transmitting whenever the corresponding output $y_i$ remains in a small neighborhood of a constant for all future times, with $i \in \{1, \dots, N\}$.
\subsection{Completeness of maximal solutions}\label{Completeness of maximal solutions}
The results in Theorem \ref{theoremLyapunov}, Proposition \ref{TheoremLyapunovSolution} and Theorem \ref{theoremLyapunovSolutionLinear} are valid on the domain of the solutions,
 but we did not say anything yet about completeness of maximal solutions. Extra properties on the system plant and the observer are needed for this purpose. In particular, we assume that system~\eqref{eq:system} is forward complete and observer \eqref{eq:observer} has the unboundeness observability property with respect to output $\hat{x}$ \cite{angeli1999forward}, as formalized in the next assumption. 
\begin{ass}\label{forwardCompleteAssumption}
	The following hold.
	\begin{enumerate}
		\item[(i)] 	For any initial condition $x_0$ in $\R^{n_x}$ and any input in $ \mathcal{L}_\mathcal{W}$, 
		the maximal solution to \eqref{eq:system} is complete. 
		\item[(ii)] For any input $u \in \mathcal{L_{\mathcal{U}}}$ and $y \in \mathcal{L}_{\R^{n_y}} $, any maximal solution $z$ to system~\eqref{eq:observer} defined on $[0, t^\star)$  with $t^\star < \infty$ satisfies $\limsup_{t \to t^\star} |\hat{x}(t)| = \infty$. \hfill $\Box$ 
	\end{enumerate}
\end{ass}
We are now ready to prove the completeness of maximal solutions of system~\eqref{eq:HybridSystem}-\eqref{eq:jumpMap_i_Esplicit}. 

\begin{thm}\label{TheoremCompletenessOfMaximalSolutions}
Under Assumptions~\ref{assumption 1},~\ref{ISSassumption} and~\ref{forwardCompleteAssumption}, any maximal solution to system~\eqref{eq:HybridSystem}-\eqref{eq:jumpMap_i_Esplicit} is complete. \hfill $\Box$
\end{thm}

\noindent\textbf{Proof:} 
We exploit \cite[Proposition 6]{heemels2021hybrid}.
Let $w \in \mathcal{L_{\mathcal{W}}}$ and $q$ be a maximal solution to \eqref{eq:HybridSystem}-\eqref{eq:jumpMap_i_Esplicit} with $w$ as input. We denote, for the sake of convenience, $\xi:= q(0,0) \in \mathcal{Q}$. By definition of $\mathcal{C}$ and $\mathcal{D}$ in \eqref{eq:FlowSet}-\eqref{eq:JumpSet_i}, $\xi \in \mathcal{C} \cup \mathcal{D}$. Suppose $\xi \in \mathcal{C} \setminus \mathcal{D}$, we want to prove that $q$ is not trivial. Since $F$ is continuous and $w \in \mathcal{L_{\mathcal{W}}}$, from \cite[Proposition S1]{cortes2008discontinuous} there exist $\epsilon > 0$ and an absolutely continuous function $\mathfrak{z}:[0, \epsilon] \rightarrow \mathcal{Q}$ such that $\mathfrak{z}(0) = \xi$, $\dot{\mathfrak{z}}(t) = F(\mathfrak{z}(t), w(t))$ for almost all $t \in [0, \epsilon]$.
We now write $\mathfrak{z} = (\mathfrak{z}_x, \mathfrak{z}_z, \mathfrak{z}_e, \mathfrak{z}_\eta)$ where $\mathfrak{z}_e = (\mathfrak{z}_{e_1}, \dots, \mathfrak{z}_{e_N})$ and $\mathfrak{z}_\eta = (\mathfrak{z}_{\eta_1}, \dots, \mathfrak{z}_{\eta_N})$. By the definition of $F$, $\mathfrak{z}_\eta(t) \geq 0$ for any $t \in [0, \epsilon]$. Moreover, since $\xi \in \mathcal{C} \setminus \mathcal{D}$, $\mathfrak{z}(0) = \xi$ and $\mathfrak{z}$ is (absolutely) continuous, there exists $\epsilon' \in (0, \epsilon]$ such that, for any $i \in \{1, \dots, N\}$, $\gamma_i(|\mathfrak{z}_{e_i}(t)|) \leq \sigma_i\alpha_i(\mathfrak{z}_{\eta_i}(t)) + \varepsilon_i$ for almost all $t \in [0, \epsilon']$. Consequently, $\mathfrak{z}(t) \in \mathcal{C}$ for almost all $t \in [0, \epsilon']$. We have proved that the viability condition in \cite[Proposition 6]{heemels2021hybrid} holds, which implies that $q$ is non-trivial. 

To prove that $q$ is complete, we need to exclude items (\textit{b}) and (\textit{c}) in \cite[Proposition 6]{heemels2021hybrid}. Item (\textit{c}) cannot occur because $G(\mathcal{D}) \subset \mathcal{C} \cup \mathcal{D}$ and the jump set imposes no condition on $w$.
On the other hand, to exclude item (\textit{b}), $q$ must not blow up in finite time. Hence, each component of $q$ must not blow up in finite time.
Let $q = (x, z, e, \eta)$.  By Assumption~\ref{forwardCompleteAssumption}, we have that $x$ cannot blow up in finite time. Moreover, $z$ cannot do so as well in view of Proposition \ref{TheoremLyapunovSolution} and item (\textit{ii}) of Assumption \ref{forwardCompleteAssumption}. 
In addition, $e$ cannot blow up in finite time by its definition and $\eta_i$ cannot in view of its dynamics \eqref{eq: etaEquation} and because $e_i$ does not, for all $i \in \{1, \dots, N\}$. 
Hence, item (\textit{b}) in \cite[Proposition 6, item (\textit{ii})]{heemels2021hybrid} cannot occur. Consequently, we conclude that any maximal solution to \eqref{eq:HybridSystem}-\eqref{eq:jumpMap_i_Esplicit} is complete. 
\hfill $\blacksquare$

\subsection{Minimum individual inter-event time}\label{IET}
To exclude the Zeno phenomena, in this section we guarantee the existence of a strictly positive minimum time between any two transmissions of each sensor node, which is an important requirement that is needed in practical applications. For this purpose, we adopt a mild boundedness condition on plant \eqref{eq:system}.
As this property is satisfied for each sensor node, and not for the overall system, it is an \textit{individual inter-event time} property, as in \cite[Definition 3]{scheres2020event}.  Indeed, simultaneous or arbitrarily close in time transmissions performed by different sensor nodes are allowed, which cannot be avoided due to the decentralized nature of the setting, see Fig.~\ref{Fig:blockDiagramDistributed}. 

We define, like in \cite{scheres2020event}, the set of hybrid times at which a jump occurs due to a transmission of sensor $i$ for $i \in \{1, \dots, N\}$, as 
\begin{equation}
	\begin{aligned}
		\mathcal{T}_i(q):= \{&(t,j) \in \dom q :q(t,j) \in \mathcal{D}_i \textnormal{ and } \\
		&q(t,j+1) \in G_i(q(t,j))\}.
	\end{aligned}
	\label{eq:Ti set}
\end{equation}

From the definition of $\mathcal{C}_i$ and $\mathcal{D}_i$ in \eqref{eq:FlowSet_i} and \eqref{eq:JumpSet_i}, we see that the time between two consecutive transmissions of a specific sensor $i$ is lower-bounded by the time it takes for $|e_i|$ to grow from $0$, which is the value after a jump due to sensor $i$, according to \eqref{eq:jumpMap_i_Esplicit}, to at least $\gamma_i^{-1}(\varepsilon_i)$. To prove that this time is lower-bounded by a strictly positive constant, we want to exploit the fact that the time derivative of $e_i$ is bounded. For this purpose, recalling that from \eqref{eq:samplingErrorContinuous} we have $ \dot{e}_i = g_i(x,u,v)= g_i(x,w)= -\frac{\partial h_i(x)}{\partial x}f_p(x,w)$, we define the following set, for any given $\mathcal{E} > 0$,
\begin{equation}
	\begin{aligned}
	\mathcal{S}_\mathcal{E} := \bigg\{(q,w) \in \mathcal{Q} \times \mathcal{W} :\left|\frac{\partial h_i(x)}{\partial x}f_p(x,w)\right| \leq \mathcal{E}&, \\ 
	\forall i \in \{1, \dots, N\} \bigg\}&,
	\label{eq:SM set}
	\end{aligned}
\end{equation}
Note that, we can take the same $\mathcal{E}$ for all $i \in \{1, \dots, N\}$. Indeed, if this is not the case and the set $\mathcal{S}_\mathcal{E}$ in \eqref{eq:SM set} is defined with arbitrarily (large) constants $\mathcal{E}_i$, which can be different for $i \in \{1, \dots, N\}$, we can always take $\mathcal{E}:= \max_{i \in \{1, \dots, N\}}\mathcal{E}_i$, and obtain \eqref{eq:SM set}. We now restrict the flow and jump sets in \eqref{eq:FlowSet}-\eqref{eq:JumpSet_i} to obtain the following hybrid system
\begin{equation}
	\begin{aligned}
		\dot{q} &= F(q,w),&& (q,w) \in \mathcal{C}_\mathcal{E} := \left(\mathcal{C}\times \mathcal{W}\right) \cap \mathcal{S}_{\mathcal{E}} \\
		q^{+} &\in G(q),   && (q,w) \in \mathcal{D}_\mathcal{E} := \left(\mathcal{D} \times \mathcal{W}\right) \cap \mathcal{S}_{\mathcal{E}}.\\
	\end{aligned}
	\label{eq:hybridSystemCompactNew}
\end{equation}
With the sets $\mathcal{C}_\mathcal{E}$ and $\mathcal{D}_\mathcal{E}$, we essentially only consider solutions to  system \eqref{eq:HybridSystem} 
such that the norm of the derivative of $e_i$ is bounded. Hence, Theorem~\ref{theoremLyapunov}, Proposition~\ref{TheoremLyapunovSolution} and Theorem~\ref{theoremLyapunovSolutionLinear} apply to system \eqref{eq:hybridSystemCompactNew}. 
It is important to notice that the constraint \eqref{eq:SM set} does not need to be implemented in the triggering rule: it is only used here for analysis purposes. 
Moreover, this constraint is always verified as long as the solution to plant \eqref{eq:system} evolves in a compact set, which is usually the case in practical applications.

In the next theorem we prove the existence of a strictly positive individual minimum inter-event time \cite[Definition 3]{scheres2020event} between any two consecutive transmissions of any sensor node for system \eqref{eq:hybridSystemCompactNew}. 

\begin{thm}
	Consider system \eqref{eq:hybridSystemCompactNew} with $\mathcal{E} > 0 $ under Assumptions~\ref{assumption 1}-\ref{ISSassumption}. Then, for any input $w \in \mathcal{L_W}$, any solution $q$ has an individual minimum inter-event time, in the sense that 
	for any $i \in \{1, \dots, N\}$ and any $(t,j), (t',j') \in \mathcal{T}_i(q)$, 
	\begin{equation}
		t + j < t' + j' \implies t'-t \geq \tau_i
	\end{equation} 
	with $\displaystyle \tau_i:= \frac{\gamma_i^{-1}(\varepsilon_i)}{\mathcal{E}}$, for all $i \in \{1,\dots, N\}$.
	As a consequence, for any input $w \in \mathcal{L_W}$, any solution $q$ to \eqref{eq:hybridSystemCompactNew} has an average dwell-time, in the sense that, for any $(t,j)$, $(t',j') \in \dom q$ with $t + j \leq t'+j'$,
		$j - j' \leq \frac{1}{\tau}(t-t') + N$
	holds with $\tau:= \frac{1}{N}\min\{\tau_1, \dots, \tau_N\}$. 
	%
	\hfill $\Box$
	\label{theoremIndividualInterEventTime} 
\end{thm}
\noindent\textbf{Proof:}
Let $w \in \mathcal{L_W}$ and $q$ be a solution to system \eqref{eq:hybridSystemCompactNew}. Pick any $(t,j) \in \dom q$ and let $0 = t_0 \leq t_1 \leq \dots \leq t_{j+1} = t$ satisfy $\dom q \cap ([0,t]\times \{0,1,\dots,j\}) = \bigcup_{k=0}^j [t_k, t_{k+1}] \times \{k\}$. For each $k \in \{0,\dots,j\}$ and almost all $s \in [t_k, t_{k+1}]$, $(q(s,k), w(s,k)) \in \mathcal{C}_\mathcal{E}$. Then, for almost all $s \in [t_k,t_{k+1}]$, from \eqref{eq:samplingErrorContinuous} and \eqref{eq:hybridSystemCompactNew}, $(q(s,k), w(s,k)) \in \mathcal{C}_\mathcal{E} = \left(\mathcal{C} \times \mathcal{W}\right) \cap \mathcal{S}_\mathcal{E}$ 
and, in view of \eqref{eq:SM set}, 
\begin{equation}
	\frac{d}{ds}|e_i|= \left|\frac{\partial h_i(x)}{\partial x}f_p(x,w)\right| \leq \mathcal{E},
	\label{eq:ei_bounded}
\end{equation}
for all $i \in \{1, \dots, N\}$. 
Let  $i \in \{1, \dots, N\}$, from \eqref{eq:jumpMap_i_Esplicit}, when $(t_k,k) \notin \mathcal{T}_i(q)$, $e_i(t_{k+1}, k+1) = e_i(t_k,k)$. Conversely, when $(t_k,k) \in \mathcal{T}_i(q)$, $e_i(t_{k+1}, k+1) = 0$. 

Let $(t_k,k) \in \mathcal{T}_i (q)$ and $t_{k'}':= \inf\big\{t \geq t_k: \left| e_i(t, k') \right|$ $\displaystyle = \gamma_i^{-1}(\varepsilon_i) \textnormal{ with } k' \geq k \textnormal{ such that } (t,k') \in \dom q\big\}$. Note that $t_{k'}'$ is not necessary the next time after $t_k$ at which sensor node $i$ generates a transmission, and that, between $t_k$ and $t_{k'}'$ only jumps, which are not due to sensor node~$i$, may occur. Consider that there are $n \in \Zo$ of these jumps. Note that $n$ is finite because of \eqref{eq:ei_bounded} and because the sampled induced errors $e_i$ are reset to $0$ after a jump, according to \eqref{eq:jumpMap_i_Esplicit}. From \eqref{eq:ei_bounded}, we have that for all $m \in [0, n-1]$ and almost all $ s\in [t_{k+m}, t_{k+m+1}]$,

\begin{equation}
	\frac{d}{ds}|e_i(s, \cdot)| \leq \mathcal{E}.
\end{equation}
Integrating this equation and applying the comparison principle \cite[Lemma 3.4]{khalil2002nonlinear}, we obtain, for all $m \in [0, n-1]$ and almost all $ s\in [t_{k+m}, t_{k+m+1}]$, 
	$|e_i(s,k+m)| \leq |e_i(t_{k+m}, k+m)| + \mathcal{E}(s - t_{k+m}). $
Similarly, for all $s \in [t_{k+n}, t_{k}']$, 
	$|e_i(s,k+n)| \leq |e_i(t_{k+n}, k+n)| + \mathcal{E}(s - t_{k+n}). $
Moreover, recalling that when $(t_k,k) \notin \mathcal{T}_i(q)$, $e_i(t_{k+1}, k+1) = e_i(t_k,k)$, 
we obtain that, for all $s \in [t_k, t_{k}']$
\begin{equation}
	|e_i(s,k')| \leq |e_i(t_{k}, k)| + \mathcal{E}(s - t_{k}), 
	\label{eq:eqProofMIET4}
\end{equation}
for $k' \in [k, k+n]$, such that $(s, k') \in \dom q$.
Moreover, since $(t_k,k) \in \mathcal{T}_i (q)$, $e_i(t_k,k) = 0$ and \eqref{eq:eqProofMIET4} becomes
\begin{equation}
	|e_i(s,k')| \leq  \mathcal{E}(s - t_{k}), \ \ \forall s \in [t_k, t_{k}'].
	\label{eq:eqProofMIET5}
\end{equation}


As a consequence, the time it takes for $ s\mapsto \mathcal{E}(s-t_k)$  to grow from $0$ to $\displaystyle \gamma_i^{-1}(\varepsilon_i)$ is $ \tau_i= \frac{\gamma_i^{-1}(\varepsilon_i)}{\mathcal{E}} > 0$, for all $i \in \{1, \dots, N\}$ and it lower-bounds $t_{k}'- t_k$ in view of \eqref{eq:eqProofMIET5}. 
Let $w \in \mathcal{L_{\mathcal{W}}}$ and $q$ be a solution to system \eqref{eq:hybridSystemCompactNew}. Pick any $(t,j), (t',j') \in \dom q$ such that $t+j \leq t'+j'$. For any $i \in \{1, \dots, N\}$, denote with $n_i(t,t')$ the number of transmission of node $i$ that occur between $(t,j)$ and $(t',j')$. In view of the above developments, we have that $n_i(t,t') \leq \frac{t'-t}{\tau_i} +1$. Noting that $\sum\limits_{i=1}^{N}n_i(t,t') = j'-j$, we have $j'-j \leq \sum\limits_{i=1}^{N}(\frac{t'-t}{\tau_i} +1)$. Using $\tau= \frac{1}{N}\min\{\tau_1, \dots, \tau_N\}$ and we obtain $j'-j \leq \frac{1}{\tau}(t'-t) + N$, which concludes the proof. 
%
\hfill $\blacksquare$


The event-triggered observer presented in this paper guarantees a strictly positive individual minimum inter-event time between  transmissions according to Theorem~\ref{theoremIndividualInterEventTime}. Therefore, the time between any two consecutive transmissions of sensor $i$ is always greater or equal than the strictly positive constant $\tau_i$
, which can be arbitrarily tuned using the design parameter $\varepsilon_i$. However, the larger $\tau_i$ is desired or needed for a practical application, the larger $\varepsilon_i$ has to be chosen and consequently, $\nu$ in Theorem~\ref{theoremLyapunov} increases. 
Note that to guarantee the individual minimum inter-transmissions time we do not need Assumption~\ref{forwardCompleteAssumption}.



\subsection{A condition for transmissions to stop}\label{stopTransmissionSubsection}
The proposed triggering rule stops the transmissions of sensor $i$ when the sampling-induced error $e_i$ becomes and remains small enough, with $i \in\ \{1, \dots, N\}$. Moreover, if the sampling-induced errors of all sensors become and remain small enough, no transmissions occurs anymore. This is formalized in the next lemma. 

\begin{lem}
	Under Assumptions~\ref{assumption 1}-\ref{ISSassumption}, consider system \eqref{eq:HybridSystem}, given a solution $q$ with input $w \in \mathcal{L_W}$, if there exists $(t,j) \in \dom q$ 
	such that
	\begin{equation}
		|e_i(t',j')| < \gamma_i^{-1}(\varepsilon_i)
		\label{eq:stopTransmission_i}
	\end{equation}
	for all $(t', j') \in \dom q$ with $t'+j' \geq t + j$, $i \in \{1,\dots, N\}$, then $\sup_j\mathcal{T}_i(q)  < \infty$. 
	In addition, if \eqref{eq:stopTransmission_i} holds for all $i \in \{1, \dots, N\}$, then $\sup_j \dom q < \infty$.
	\label{Lemma_stopTransmissions}\hfill $\Box$ 
\end{lem}
\noindent\textbf{Proof:}
Let $q$ be a solution  to system \eqref{eq:HybridSystem} with input $w \in \mathcal{L_W}$. The condition $|e_i(t',j')| < \gamma_i^{-1}(\varepsilon_i)$ for all $(t',j') \in \dom q$ with $t'+j' \geq t + j$ in \eqref{eq:stopTransmission_i} implies that $\gamma_i(|e_i(t',j')|) < \gamma_i(\gamma_i^{-1}(\varepsilon_i)) = \varepsilon_i \leq \sigma_i\alpha_i(\eta_i) + \varepsilon_i$ for all $(t',j') \geq (t,j)$ with $(t',j') \in \dom q$. Therefore, no jumps due to sensor $i$ occurs after $(t,j)$. Hence, $\sup_j \mathcal{T}_i(q) < \infty$. 

Moreover, if the condition  $|e_i(t',j')| < \gamma_i^{-1}(\varepsilon_i)$ is satisfied for all $i \in \{1, \dots N\}$, then, from the first part of this proof we have $\sup_j \mathcal{T}_i(q) < \infty$ for all $i \in \{1, \dots, N\}$. As a consequence, $\max_{i \in \{1, \dots, N\} } \{\sup_j \mathcal{T}_i(q)\} < \infty$. From \eqref{eq:JumpSet}, \eqref{eq:JumpSet_i}, a jump can occur only when one or more sensors need to transmit, therefore, from \eqref{eq:Ti set}, $\sup_j \dom q = \max_{i \in \{1, \dots, N\} }\{\sup_j\mathcal{T}_i(q)\} < \infty$. 
\hfill $\blacksquare$

Condition \eqref{eq:stopTransmission_i} occurs when the output $y_i$, $i \in \{1, \dots, N\}$, remains in a small neighborhood of a constant for all positive times for instance. Indeed, when, for some constant $y_i^* \in \R^{n_{y_i}}$, the output $y_i$ satisfies $\displaystyle |y_i(t)-y_i^*|<\frac{1}{2} \gamma_i^{-1}(\varepsilon_i)$ for all $t \geq T$ for some $T\geq 0$, then for any solution $q$ to \eqref{eq:HybridSystem} and any $(t_{j_i}, j_i), (t,j) \in \dom q$, with $(t_{j_i}, j_i-1) \in \mathcal{T}_i(q)$ and $t_{j_i} \geq T$, $t \geq t_{j_i}$, $j \geq j_i$ and $\displaystyle |e_i(t,j)| = |y_i(t_{j_i}, j_i) -y_i(t,j)| = |y_i(t_{j_i}, j_i) -y_i^* + y_i^* -y_i(t,j)|\leq |y_i(t_{j_i}, j_i) -y_i^*| + |y_i^* -y_i(t,j)|<2\frac{1}{2}\gamma_i^{-1}(\varepsilon_i)$ and \eqref{eq:stopTransmission_i} holds. Moreover, sensor $i$ automatically starts transmitting again if condition \eqref{eq:stopTransmission_i} is no longer satisfied. This is a clear advantage over time-triggered strategies, where output $y_i$ is always transmitted, even if its information is not needed to perform the estimation; see \cite[Figure 3]{petri2021Event} for an illustration. 
It is worth noting that Lemma~\ref{Lemma_stopTransmissions} applies to system \eqref{eq:HybridSystem}, and not only to system \eqref{eq:hybridSystemCompactNew}. Therefore, it is not necessary restrict the flow and jump sets with the $\mathcal{S}_\mathcal{E}$ set in \eqref{eq:SM set}. 
Moreover, as for Theorem~\ref{theoremIndividualInterEventTime}, Assumption~\ref{forwardCompleteAssumption} is not needed for this result. 
%
%
%
%
\section{Extensions}\label{discussionSection}
In this section, we discuss generalizations and extensions of the results presented so far. In Section~\ref{generalTriggeringCondition}, we explain how the triggering condition can be generalized, 
while in Section~\ref{MeasurementNoise} we discuss the modifications needed in presence of measurement noise. Finally, in Section~\ref{TriggeredInput} we consider the case when the input $u$ is sampled and transmitted to the observer via a digital network and we propose a triggering condition for $u$, which is compatible with the previous results.  

\subsection{Generalized triggering conditions}\label{generalTriggeringCondition}
The $\eta_i$-system and the triggering rule in~\eqref{eq: etaEquation} and~\eqref{eq:triggeringRule} are special cases of a more general $\eta_i$-system and a more general triggering rule that guarantee the stability results. Indeed, we can design the auxiliary scalar variable $\eta_i$ with the following dynamics instead of~\eqref{eq: etaEquation}, for all $i \in \{1,\dots, N\}$,
\begin{equation}
	\dot{\eta}_i := -\widecheck{\alpha}_i(\eta_i) + \widecheck{\gamma}_i(|e_i|),
	\label{eq:etaDotRemarkGeneralTriggering}
\end{equation}
with any $\widecheck{\alpha}_i \in \Kinf$ and any $\widecheck{\gamma}_i \in \Kinf$. 
Regarding the triggering rule, let $\mathfrak{d}_i$ be any non-decreasing continuous function from $\R_{\geq0}$ to $\R_{\geq0}$, which can be equal to $0$ only at $0$. 
The triggering rule in \eqref{eq:triggeringRule} can then be replaced by
\begin{equation}
	\gamma_i(|e_i|) + \frac{\partial d_i(\eta_i)}{\partial \eta_i} \widecheck{\gamma}_i (|e_i|) \leq    \sigma_i \frac{\partial d_i(\eta_i)}{\partial \eta_i} \widecheck{\alpha}_i (\eta_i) + \varepsilon_i,
	\label{eq:triggeringRuleRemarkGeneralTriggering}
\end{equation}
where $d_i \in \Kinf$ is defined as $ d_i(s):= \int_{0}^{s} \mathfrak{d}_i(\tau) d\tau$ for all $s \geq 0$ and $\sigma_i \in (0,1)$ for all $i \in \{1, \dots, N\}$. 
We can then follow the same lines as in Section~\ref{StabilityProofsSubsection} to obtain Lyapunov and stability results like in Sections~\ref{LyapunovPropertiesSubsection}-\ref{StabilitySolutionsSubsection}.

\subsection{Additive measurement noise}\label{MeasurementNoise}
In the case where the system output is affected by additive measurement noise, system~\eqref{eq:system} becomes
\begin{equation}
	\begin{aligned}
		\dot x  &=  f_p(x,u,v)\\
		\tilde{y}  &=  h(x) + m,
	\end{aligned}
	\label{eq:systemWithMeasurementNoise}
\end{equation}
with $m \in \mathcal{L}_{\mathcal{M}}$, where $\mathcal{M}:= \mathcal{M}_1 \times \dots \times \mathcal{M}_N \subseteq \R^{n_{y_1}} \times \dots \times \R^{n_{y_N}}$. 
The output measured by sensor $i$, with $i \in \{1, \dots, N\}$ is 
\begin{equation}
	\tilde{y}_i = y_i + m_i
\end{equation}
where $m_i \in \mathcal{L}_{\mathcal{M}_i}$ is the measurement noise of sensor $i$. 
We assume that we know a bound on the $\mathcal{L}_\infty$-norm of the measurement noise. Therefore, the set $\mathcal{M}_i$ is defined as 
\begin{equation}
	\mathcal{M}_i := \{\widecheck{m}_i \in \R^{n_{y_i}} : |\widecheck{m}_i| \leq \mathfrak{m}_i\} 
\end{equation}
for some $\mathfrak{m}_i \in \R_{\geq0}$. 
Consequently, the observer does not know the real output $y_i$, but its sampled noisy version, due to the network, $\tilde{\bar{y}}_i := \bar{y}_i + \bar{m}_i$, where $\bar{m}_i$ is the networked version of the measurement noise $m_i$
, with $i \in \{1, \dots, N\}$. Due to the measurement noise, sensor $i$ does not know the network-induced error $e_i$, but only $\tilde{e}_i$, which is the network-induced error of sensor $i$ in presence of noise, which is defined following \cite{scheres2020event}, 
\begin{equation}
	\tilde{e}_i := \tilde{\bar{y}}_i - \tilde{y}_i = \bar{y}_i + \bar{m}_i -y_i -m_i = e_i + \bar{m}_i -m_i
\end{equation}
for all $i \in \{1, \dots, N\}$. As a consequence, the triggering rule cannot rely on $e_i$, and sensor $i$ needs to decide when the measured output $\tilde{y}_i$ has to be transmitted to the observer based on $\tilde{e}_i$. We therefore replace the dynamic of $\eta_i$ in \eqref{eq: etaEquation} by $\dot{\tilde{\eta}}_i = -\alpha_i(\tilde{\eta}_i) + c_i\gamma_i(|\tilde{e}_i|)$ and the triggering rule in \eqref{eq:triggeringRule} by $\gamma_i(|\tilde{e}_i|) \geq \sigma_i\alpha_i(\tilde{\eta}_i) + \varepsilon_i$, 
for all $i \in \{1, \dots, N\}$. 
We can then follow similar lines as in \cite{scheres2020event} to guarantee a practical input-to-state stability property for the estimation error system and a 
semi-global individual  minimum inter-event time.  We just need to select $\varepsilon_i > \gamma_i(2\mathfrak{m}_i)$, for all $i \in \{1, \dots, N\}$ and then all the previous results hold. 
Note that, since, in presence of measurement noise we have a lower-bound on $\varepsilon_i$, for all $i \in \{1, \dots, N\}$, we cannot select $\nu$ arbitrary small, as in Theorem \ref{theoremLyapunov}.

\subsection{Triggering the input $u$} \label{TriggeredInput} 
When the input $u$ to \eqref{eq:system} is communicated to the observer over a digital network, Assumption~\ref{assumption 1} does not hold. We explain how to define a triggering rule for $u$ in this case so that the previous results apply \textit{mutatis mutandis}.

Let $\bar{u}$ be the networked version of $u$ available to the observer. Between two successive transmission instants, using zero-order-hold device we have $\dot{\bar{u}} = 0$, 
and when the input is sent, $	u^+ = u.$
We define the input network-induced error $e_u$ as $e_u:= \bar{u} - u$
and the observer equations in \eqref{eq:observerNew} becomes
\begin{equation}
	\begin{aligned}
		\dot{ z}  &= f_o(z,\bar{u},\bar{y},\hat{y}) = f_o(z,u+e_u, y + e,\hat{y}), \\
		\hat{x} &= \psi(z), \qquad
		\hat y  = h(\hat{x}).
	\end{aligned}
	\label{eq:observerNewInputTriggered}
\end{equation}
In this new setting, where also the input is sampled, Assumption~\ref{ISSassumption} needs to be modified so that an input-to-state stability property holds also with respect to the input sampled-induced error $e_u$. 
\begin{ass}
	There exist $\underline{\alpha}$, $\overline{\alpha}$, $\alpha$, $\gamma_1, \dots, \gamma_N $, $  \theta$, $\gamma_u$ $\in \Kinf$, $V: \R^{n_x} \times \R^{n_z} \rightarrow \R_{\geq0}$ continuously differentiable, such that for all $x \in \R^{n_x}$, $z \in \R^{n_z}$, $u \in \R^{n_u}$, $e \in \R^{n_y}$, $\hat{y} \in \R^{n_y}$, $v \in \R^{n_v}$, $e_u \in \R^{n_u}$, \eqref{eq:ISSassumptionSandwichBound} holds and 
	\begin{equation}
		\underline{\alpha}(|x-\psi(z)|) \leq V(x,z) \leq \overline{\alpha}(|\psi^{-R}(x) - z|)	
		\label{eq:ISSassumptionSandwichBoundInputTriggered}	
	\end{equation}
	\begin{equation}
		\begin{array}{l}
		\left\langle \nabla V(x,z), (f_p(x,u,v),f_o(z,u,y + e, \hat{y})) \right\rangle  \leq  \\[.5em]
		\quad -\alpha(V(x,z)) + \sum\limits_{i=1}^{N} \gamma_i(|e_i|) + \theta(|v|) + \gamma_u(|e_u|).
	\end{array}
		\label{eq:ISSassumptionDerivativeInputTriggered}
	\end{equation}\hfill $\Box$
%
	\label{ISSassumptionInputTriggered}
\end{ass}
	For many classes of observers in the literature, if the observer is input-to-state stable with respect to $v$, then it is also input-to-state stable with respect to $e_u$, see \cite{astolfi2021stubborn} for more details.

Based on Assumption~\ref{ISSassumptionInputTriggered}, we can design the triggering rule for the input similarly to the triggering rule designed in \eqref{eq:triggeringRule} for the output $y_i$, with $i \in \{1, \dots, N\}$. In particular, let  $\eta_u$ be an auxiliary scalar variable, whose equations during flows and jumps are, respectively, 
\begin{equation}
	\begin{aligned}
		\dot{{\eta}}_u = & - \alpha_u(\eta_u) + c_u\gamma_u(|e_u|)=: \ell_u(\eta_u, e_u)\\
		\eta_u^+ = & \ b_u \eta_u 
	\end{aligned}
	\label{eq:etaEquationTriggeredInput}
\end{equation}
where $\gamma_u$ comes from Assumption~\ref{ISSassumptionInputTriggered} and $\alpha_u \in \Kinf$, $ c_u \geq 0$ and $b_u \in [0,1]$ are design function and parameters. An input data is transmitted to the observer when the condition 
\begin{equation}
	\gamma_u(|e_u|) \geq \sigma_u \alpha_u(\eta_u) + \varepsilon_u
	\label{eq:inputTriggeringRule}
\end{equation}
is satisfied, where  $\sigma_u \geq 0$ and $\varepsilon_u > 0$ are design parameters. As for the output triggering rule, parameter $\varepsilon_u$ is needed to avoid the Zeno phenomena. 
In this new setting, all previous stability results apply similarly. Moreover, to have an individual minimum inter-event time a sufficient condition is that the input $u$ is continuously differentiable and $|\dot{u}| \leq \mathcal{E}_u$, where $\mathcal{E}_u$ is any positive constant. 
\section{Numerical case study}\label{Example}
We design the event-triggered observer presented in this paper to a flexible joint robotic arm \cite{raff2008observer}. For this application, our framework is relevant in scenarios where the observer is not co-located with the robotic arm and communicates with it through a digital network. In this case study, we consider two sensor nodes, but the results would also be relevant if we would have only one node. The system model is described by
\begin{equation}
	\begin{aligned}
	\dot{x} &= Ax + Bu + G\sigma(Hx) +v\\
	y &= Cx + m,
	\end{aligned}
\end{equation}
where the system state that need to be estimated is $x:=(x_1, x_2, x_3, x_4)$, while the measured output $y$ is defined as $y := (y_1, y_2) = (x_1, x_2)$. The system matrices are
\footnotesize \begin{equation}
	\begin{array}{l}
	A = \begin{bmatrix}
		0 & 1 & 0 & 0\\
		-48.6 & -1.25 & 48.6 & 0\\
		0 & 0 & 0 & 1\\
		19.5 & 0 & -19.5 & 0
	\end{bmatrix}, \quad
	B  = \begin{bmatrix}
	0 \\
	21.6\\
	0 \\
	0
\end{bmatrix}, \\[0.5em]
G = \begin{bmatrix}
	0 \\
	0\\
	0 \\
	-1
\end{bmatrix}, \quad
H^\top = \begin{bmatrix}
	0 \\
	0\\
	1 \\
	0
\end{bmatrix}, \quad
C^\top = \begin{bmatrix}
	1 &0 \\
	0 &1 \\
	0 &0 \\
	0 &0
\end{bmatrix},\\[0.5em]
\end{array}
\end{equation} \normalsize
and $\sigma(Hx) = 3.3 \sin(x_3)$ for any $x \in \R^4$.
As in \cite{raff2008observer}, we assume that the input is $u(t) = \sin(t)$ for all $t \in \R_{\geq0}$. Moreover, we consider the disturbance input $\displaystyle v(t) = 0.02(0,1,0,1)\sin(0.4t)$ 
for all $t \in \R_{\geq0}$  and the measurement noise $\displaystyle m(t) = 0.01 (0,1)\sin(0.3t)$
 for all $t \in \R_{\geq0}$.
We design a continuous-time observer 
\begin{equation}
	\begin{aligned}
		\dot{\hat{x}} &= A\hat{x} + Bu + G\sigma(H\hat{x}) + L(y -\hat{y})\\
		\hat{y} &= C\hat{x},
	\end{aligned}
\label{eq:observerExample}
\end{equation}
where $L \in \R^{4\times 2}$ is the observer gain that is designed following a polytopic approach \cite{zemouche2008observers}. To do so, we solve the 
linear matrix inequalities $PA -WC + PG_i +G_i^\top P +A^\top P -C^\top W^\top \leq -Q, \ \  i \in \{1,2\}$, with $P \in \R^{4 \times 4}$ symmetric positive definite and $W:= PL \in \R^{4 \times 2}$,
where \footnotesize $\displaystyle G_1:= \begin{bmatrix}
	0 & 0 & 0 & 0\\
	0 & 0 & 0 & 0\\
	0 & 0 & 0 & 0\\
	0 & 0 & 3.3 & 0\\
\end{bmatrix}, 
G_2:= \begin{bmatrix}
0 & 0 & 0 & 0\\
0 & 0 & 0 & 0\\
0 & 0 & 0 & 0\\
0 & 0 & -3.3 & 0\\
\end{bmatrix}$ \normalsize and $Q = I_4$. We obtain \footnotesize $L = \begin{bmatrix}
 0.58 &-42.96\\
 -4.67 & 2.83 \\
 3.16 & 49.25\\
 16.34 & 88.46
\end{bmatrix}$. \normalsize
Note that, observer \eqref{eq:observerExample} is in the form of \eqref{eq:observerAlly} with $z = \hat{x}$. 
Defining the Lyapunov function $V(\xi) := \xi^{\top}P\xi$ for any $\xi \in \R^4$, where $\xi:= x - \hat{x}$ is the state estimation error, Assumption~\ref{ISSassumption} is satisfied with $ \alpha(s) = \frac{\lambda_{\min}(Q) - \mathfrak{c}_v -\mathfrak{c}_1-\mathfrak{c}_2}{\lambda_{\max}(P)}s $, $\theta(s) = \frac{1}{\mathfrak{c}_v}\norm{P}^2|s|^2$, $\gamma_1(s) = \frac{1}{\mathfrak{c}_1}\norm{PL_1}^2|s|^2$ and $ \gamma_2(s) = \frac{1}{\mathfrak{c}_2}\norm{PL_2}^2|s|^2$, where $\mathfrak{c}_v, \mathfrak{c}_1, \mathfrak{c}_2$ are  parameters chosen such that $\mathfrak{c}_v>0, \mathfrak{c}_1 >0, \mathfrak{c}_2 >0$ and $\lambda_{\min}(Q) - \mathfrak{c}_v - \mathfrak{c}_1 - \mathfrak{c}_2 >0$, while $L_1$ and $L_2$ are the first and the second column of the matrix gain $L$, respectively. 

We have first simulated the event-triggered observer \eqref{eq:HybridSystem}-\eqref{eq:jumpMap_i_Esplicit} with $\sigma_1 = 600$, $\sigma_2 = 800$, $c_1 =0.001$, $c_2=0.001$, $b_1 =1$, $b_2 = 1$, $\alpha_1 (s) = a_1s$, with $a_1 =2$, $\alpha_2 (s) = a_2s$, with $a_2 =3$, $\varepsilon_1 = 10$ and $\varepsilon_2 = 10$. With this choice of parameters the conditions $\sigma_1c_1 < 1$ and $\sigma_2c_2 < 1$ are satisfied and Theorems~\ref{theoremLyapunov} and~\ref{theoremLyapunovSolutionLinear} apply. Moreover, the condition $ \left|\frac{\partial h_i(x)}{\partial x}f_p(x,w)\right| \leq \mathcal{E}$ is satisfied for $i \in \{1,2\}$, for $\mathcal{E}$ large enough and Theorem \ref{theoremIndividualInterEventTime} applies. 
Thanks to the freedom on the choice of $\gamma_i$ in Remark~\ref{remarkFreedomISSgains}, 
we do not need to use $\gamma_1$, $\gamma_2$ coming from Assumption~\ref{ISSassumption}, as explained in Section~\ref{TriggeringRuleAndHybridModel}, but we can select any $\gamma_1$, $\gamma_2$ such that $\gamma_1(s) = \mathfrak{l}_1 s^2$ and $\gamma_2(s) = \mathfrak{l}_2 s^2$, with $\mathfrak{l}_1>0$ and $\mathfrak{l}_2>0$, which are thus additional design parameters. We select $\gamma_1(s) = 5 s^2$ and $\gamma_2(s) = 5 s^2$. 

\begin{figure}
	\centering
	\includegraphics[trim= 4.7cm 7.45cm 4.7cm 7.92cm, clip, width=0.8\linewidth]{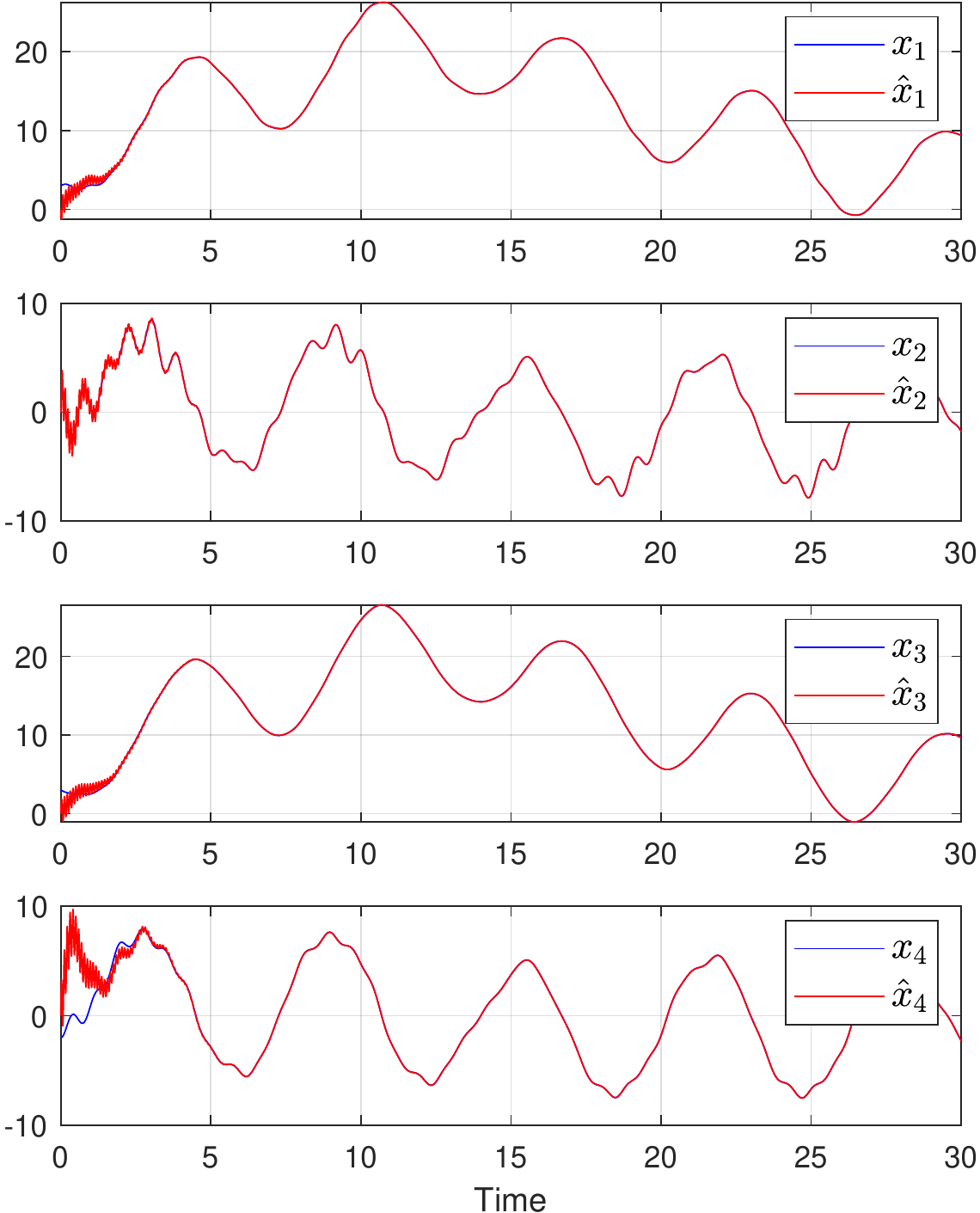}
	\caption{\normalfont{State $x$ and state estimate $\hat{x}$ 
	}}
	\label{fig:StateAndEstimate}
\end{figure}

\begin{figure}[t]
	\centering
	\includegraphics[trim= 0.3cm 11.8cm 0.4cm 11.8cm, clip, width=0.9\linewidth]{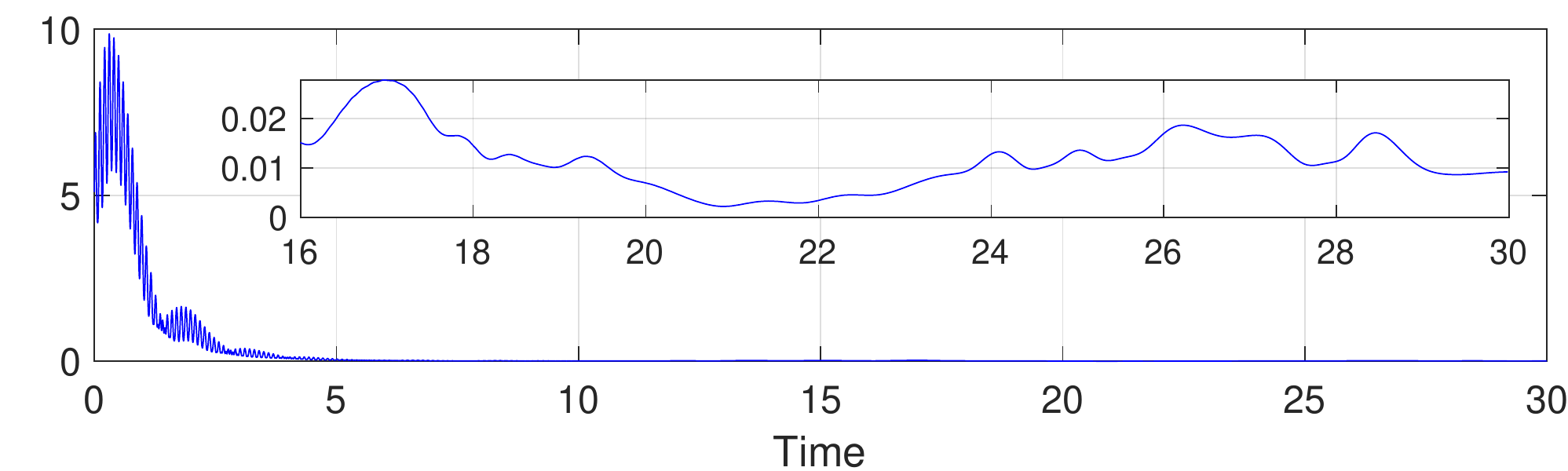}
	\caption{\normalfont{Norm of the state estimation error $|\xi|:= |x-\hat{x}|$ 
	}}
	\label{fig:NormEstimationError}
\end{figure}

\begin{figure}[t]
	\centering
	\includegraphics[trim= 4.3cm 10.35cm 4.6cm 11.1cm, clip, width=0.8\linewidth]{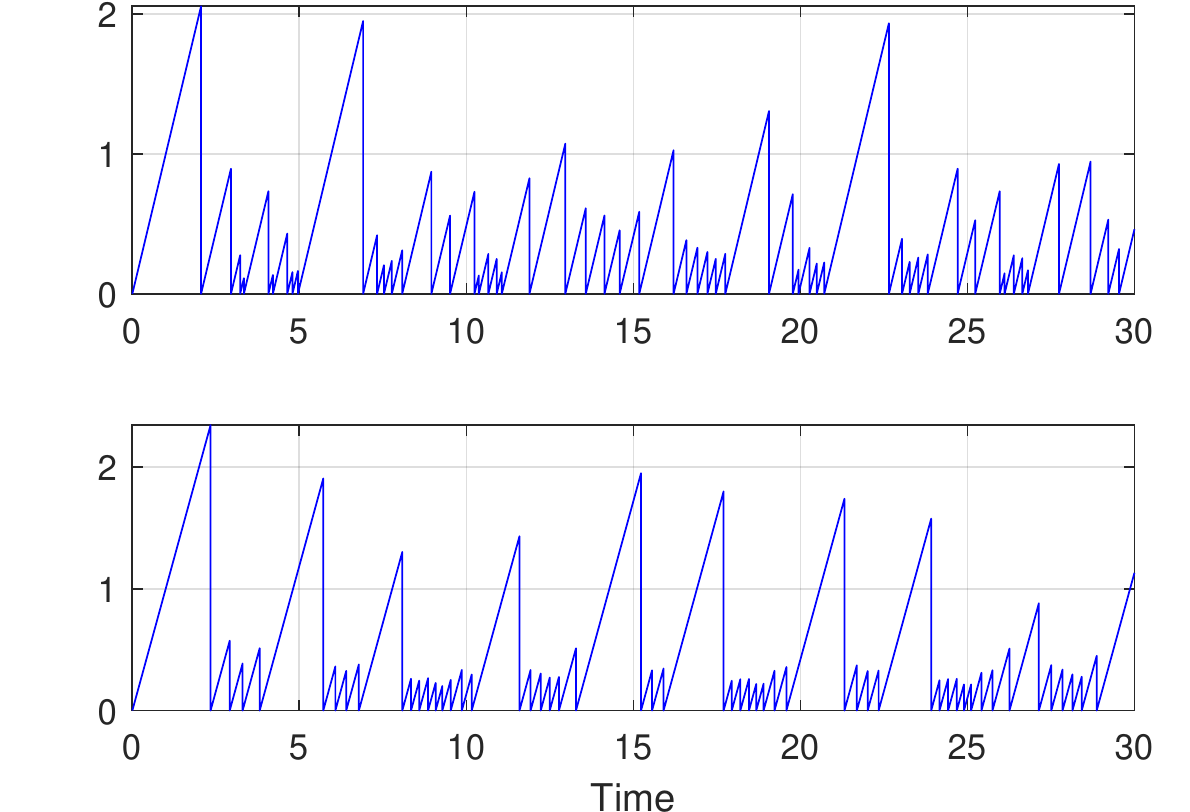}
	\caption{\normalfont{Inter-transmissions times (sensor $1$ top, sensor $2$ bottom) 
		}}
	\label{fig:InterTransmissionsTime}
\end{figure}

We have considered the following initial conditions $x(0,0) = (3,2,3,-2), \hat{x}(0,0) = (0, 0, 0, 0), e(0,0) = (0,0)$ and $\eta(0,0) = (10, 10)$. 
In Figure~\ref{fig:StateAndEstimate}, we provide the plots obtained for the plant states and its estimates, in Figure~\ref{fig:NormEstimationError} the plot related to the norm of the estimation error is shown, while in Figure~\ref{fig:InterTransmissionsTime} the inter-transmissions time is reported. From these figures, it is clear that all state estimation error practically converge. Moreover, the minimum inter-event time measured is $0.201 \, \textnormal{s}$ for sensor $1$ and $0.112\, \textnormal{s}$ for sensor~$2$.  

%

We have also analyzed the impact of the design parameters, in particular we focus on the effect of $\sigma_1$, $\sigma_2$, $\varepsilon_1$, $\varepsilon_2$, $a_1$, $a_2$, $\mathfrak{l}_1$ and $\mathfrak{l}_2$. We have run for this purpose simulations with different parameters configurations and $100$ different initial conditions for each chosen parameters configuration. In particular, $x_1(0,0)$ and $x_3(0,0)$ were selected randomly in the interval $[0,20]$, while $x_2(0,0)$ and $x_4(0,0)$ were chosen randomly in the interval $[0,10]$. The initial conditions of the observer states $\hat{x}_1(0,0), \hat{x}_2(0,0), \hat{x}_3(0,0), \hat{x}_4(0,0)$ and of the network-induced errors $e_1(0,0), e_2(0,0)$ were always selected equal to $0$, while $\eta_1(0,0) = \eta_2(0,0) = 10$ in all simulations. For all the choice of parameters, we have evaluated the number of transmissions in the (continuous) time interval $[0, 30]$ on average and the maximum ultimate bound on the state estimation error in the time interval $[20, 30]$ averaged over all simulations. The data collected are shown in Table~\ref{tab:SimulationData}. The same analysis was repeated also in the case where the system is not affected by the disturbance input $v$ and the measurement noise $m$. In Table~\ref{tab:SimulationData} the data collected in this configuration are also reported.

\begin{table*}[t]
	\caption{\normalfont{Average number of transmissions in the time interval $[0,30]$ 
			and maximum absolute value of the state estimation error $|\xi|$ for $t \in [20,30] $ 
			with different choices for $\sigma_1$, $\sigma_2$, $\varepsilon_1$, $\varepsilon_2$, $a_1$, $a_2$, $\mathfrak{l}_1$ and $\mathfrak{l}_2$, both with and without disturbance input and measurement noise.}}
	\label{tab:SimulationData}
	\begin{center}
		\scriptsize
		\begin{tabular}{cccccccc|cc|cc}
			\toprule
			$\sigma_1$& $\sigma_2$& $\varepsilon_1$& $\varepsilon_2$& $a_1$& $a_2$& $\mathfrak{l}_1$& $\mathfrak{l}_2$ & Transmissions & $|\xi|$ & Transmissions & $|\xi|$ \\
			& & & & & & & & with $v$ and $m$ & with $v$ and $m$ & without $v$ and $m$ & without $v$ and $m$\\ 
			\toprule
			$600$ & $800$ & $10$& $10$ &$2$ & $3$ & $5$ & $5$ & $163$ &$0.0236$ & $167$ & $6.32 \cdot 10^{-5}$\\
			\midrule
			$600$ & $800$ & $1$& $1$ &$2$ & $3$ & $5$ & $5$ & $497$ &$0.0235$ & $515$ & $2.13 \cdot 10^{-5}$ \\
			$600$ & $800$ & $100$& $100$ &$2$ & $3$ & $5$ & $5$ & $47$ &$0.0236$ & $49$ & $2.34 \cdot 10^{-4}$ \\
			$600$ & $800$ & $1000$& $1000$ &$2$ & $3$ & $5$ & $5$ & $10$ &$0.0234$ & $7$ & $2.63 \cdot 10^{-4}$\\
			\midrule
			$0$ & $0$ & $10$& $10$ &$2$ & $3$ & $5$ & $5$ & $452$ &$0.0238$ & $474$ & $4.02 \cdot 10^{-5}$ \\
			$300$ & $400$ & $10$& $10$ &$2$ & $3$ & $5$ & $5$ & $221$ &$0.0235$ & $214$ & $4.98 \cdot 10^{-5}$ \\
			$950$ & $950$ & $10$& $10$ &$2$ & $3$ & $5$ & $5$ & $148$ &$0.0236$ & $156$ & $7.43 \cdot 10^{-5}$\\
			\midrule
			$600$ & $800$ & $10$& $10$ &$1$ & $1.5$ & $5$ & $5$ & $126$ &$0.0238$ & $125$ & $1.14 \cdot 10^{-4}$\\
			$600$ & $800$ & $10$& $10$ &$4$ & $6$ & $5$ & $5$ & $223$ &$0.0235$ & $228$ & $6.08 \cdot 10^{-5}$\\
			$600$ & $800$ & $10$& $10$ &$10$ & $10$ & $5$ & $5$ & $267$ &$0.0234$ & $238$ & $4.01 \cdot 10^{-5}$\\
			\midrule
			$600$ & $800$ & $10$& $10$ &$2$ & $3$ & $1$ & $1$ & $55$ &$0.0236$ & $52$ & $2.01 \cdot 10^{-4}$\\
			$600$ & $800$ & $10$& $10$ &$2$ & $3$ & $10$ & $10$ & $256$ &$0.0236$ & $256$ & $2.54 \cdot 10^{-5}$\\
			$600$ & $800$ & $10$& $10$ &$2$ & $3$ & $100$ & $100$ & $922$ &$0.0236$ & $923$ & $9.88 \cdot 10^{-7}$\\
			\bottomrule
		\end{tabular} \normalsize
	\end{center}
\end{table*}

Table~\ref{tab:SimulationData} shows that choice of the design parameters impacts the average number of transmissions both when the system is affected by the additional disturbance input $v$ and measurement noise $m$ and when it is not. Moreover, data shows that the ultimate bound of the estimation error is small in all the chosen configurations and that the obtained values are not significantly affected by the choice of the parameters in presence of noise $m$ and disturbance $v$, but this is no longer true when those are absent.

\section{Conclusions}\label{Conclusions}
We have presented a decentralized event-triggered observer design for perturbed nonlinear systems. We have designed for this purpose new dynamic triggering rules for each sensor node to define the transmissions over the digital network. 
We have formally established a uniform global practical stability property for the estimation error and we guarantee the existence of a uniform, strictly positive time between any two transmissions of each sensor node. 
Moreover, the proposed triggering rule does not require significant computation capability on the smart sensor, as it only needs to run a local scalar filter. We have also shown how the triggering rule can be generalized and how to cope with measurement noise and/or sampled input. 

%
%

It would be interesting in future work to tailor the results to specific classes of systems and observers, as we did for linear time-invariant systems in \cite{petri2021Event}. Another relevant research direction would be to take into account other network effects such as delays and packet losses, by taking inspiration from e.g., \cite{dolk2017event, dolk2016output}. 

 \appendix \label{Appendix}
 \section{Technical lemmmas}
We present two technical lemmas. The first one is about the change of the supply rates and generalizes \cite[Theorem 1]{sontag1995changing}. 
 \begin{lem}
 	Let $f: \R^{n_x} \times \R^{n_{u_1}} \times \dots \times \R^{n_{u_N}} \rightarrow \R^{n_x}$, with $n_x, n_{u_1}, \dots, n_{u_N} \in \Z$. 
 	Suppose there exist $V:\R^{n_x} \rightarrow \R_{\geq0}$, with $n_x \in \Zp$ continuously differentiable,  $\underline{\alpha}_V$, $\overline{\alpha}_V$, $\alpha$, $\gamma_1, \dots, \gamma_N \in \Kinf$ such that for all $x \in \R^{n_x}$, $u_i \in \R^{n_{u_i}}$, 
 	\begin{equation}
 		\begin{aligned}
 		\underline{\alpha}_V(|x|) \leq V(x)& \leq \overline{\alpha}_V(|x|)	\\
 		\left\langle \nabla V(x), (f(x,u_1, \dots u_N)) \right\rangle&  \leq -\alpha(|x|) + \sum\limits_{i=1}^{N} \gamma_i(|u_i|).
 		\label{eq:LemmaMultipleChangeSupplyRate}
 		\end{aligned}
 	\end{equation}
 	Then, for all $i \in \{1, \dots, N\}$ and any given $\tilde{\gamma}_i \in \Kinf$ verifying $\gamma_i(r) = O(\tilde{\gamma}_i(r))$ as $r \to \infty$, there exist 
 	$\underline{\alpha}_W$, $\overline{\alpha}_W$, $\tilde{\alpha}$ and $W:\R^{n_x} \rightarrow \R_{\geq0}$ continuously differentiable such that for all $x \in \R^{n_x}$, $u_i \in \R^{n_{u_i}}$, 
 	\begin{equation}
 		\underline{\alpha}_W(|x|) \leq W(x) \leq \overline{\alpha}_W(|x|)	
 		\label{eq:LemmaMultipleChangeSupplyRate_sandwichBound_W}	
 	\end{equation}
 	\begin{equation}
 		\left\langle \nabla W(x), f(x,u_1, \dots u_N) \right\rangle  \leq -\tilde{\alpha}(|x|) + \sum\limits_{i=1}^{N} \tilde{\gamma}_i(|u_i|).
 		\label{eq:LemmaMultipleChangeSupplyRate_derivative_W}
 	\end{equation}
 	\label{LemmaMultipleChangeOfSupplyRate}\hfill $\Box$ 
 \end{lem}

 \noindent\textbf{Sketch of proof:}
  The proof follows similar steps as the proof of  \cite[Theorem 1]{sontag1995changing}.
  Let $W: = \rho \circ V$
  where  $\rho$ is a $\Kinf$-function defined as
	$\rho(s):= \int_{0}^{s} q(t) dt$,
  where $q$ is a suitably chosen smooth non-decreasing function from $[0, \infty)$ to $[0, \infty)$, which satisfies $q(t) > 0$ for $t >0$. 
  Hence, the function $W$ is continuously differentiable and positive definite by properties of $\rho$ and $V$. As a consequence, there exist $\underline{\alpha}_W \in \Kinf$ and $\overline{\alpha}_W \in \Kinf$ such that \eqref{eq:LemmaMultipleChangeSupplyRate_sandwichBound_W} is satisfied. 
  Following similar steps as in the proof of  \cite[Theorem 1]{sontag1995changing} we obtain 
		$\left\langle \nabla W(x), f(x,u_1, \dots u_N) \right\rangle \leq 
 \sum\limits_{i=1}^{N} \left[q(V(x))\left(-\frac{1}{N}\alpha(|x|) + \gamma_i(|u_i|)\right)\right]$
%
  instead of \cite[Equation (4)]{sontag1995changing} and 
   \begin{equation}
  	\begin{array}{l}
  		\left\langle \nabla W(x), f(x,u_1, \dots u_N) \right\rangle \leq\\[.5em]
  		\qquad \sum\limits_{i=1}^{N} \left[q(\vartheta_i(|u_i|))\gamma_i(|u_i|) - \frac{1}{2N}q(\underline{\alpha}_V(|x|))\alpha(|x|)\right]
  	\end{array}
  	\label{eq:multipleChangeSupplyRate:eqSketchProof3}
  \end{equation}
  instead of \cite[Equation (6)]{sontag1995changing}  with $\vartheta_i:= \overline{\alpha}_V \circ \alpha^{-1} (2N\gamma_i) \in \Kinf$, for any $i \in \{1, \dots, N\}$.
 %
  Since $ \gamma_i(|u_i|) = O(\tilde{\gamma}_i(|u_i|))$ as $|u_i| \to \infty$, for all $i \in \{1, \dots, N\}$, following the same arguments as in the proof of \cite[Theorem 1]{sontag1995changing}  we obtain that, for all $i \in \{1, \dots, N\}$, there exists $q_i$ smooth non-decreasing function such that $q_i(0) = 0$ and 
  \begin{equation}
  	q_i(\vartheta_i(|u_i|))\gamma_i(|u_i|)\leq \tilde{\gamma}_i (|u_i|). 
  	\label{eq:multipleChangeSupplyRate:eqProof6}
  \end{equation}
 Note that the condition  $q_i(0) = 0$ does not come from the proof of \cite[Theorem 1]{sontag1995changing}, but the proof applies by adding this extra condition.
  We define $\tilde{q}:= \min\{q_1, \dots, q_N\}$. Note that $\tilde{q}$ is a positive definite, non-decreasing function.
  Using \cite[Lemma 1]{kellett2014compendium} we have that there exists a function $q \in \K$, smooth on $\R_{>0}$, so that 
  $	q(s) \leq \tilde{q}(s) \leq q_i(s)$
  for all $s \geq 0$, for all $i \in \{1, \dots, N\}$. 
 Combining the last inequality with 
   \eqref{eq:multipleChangeSupplyRate:eqSketchProof3}, \eqref{eq:multipleChangeSupplyRate:eqProof6},  
   and defining $\tilde{\alpha} \in \Kinf$ as in the proof of \cite[Theorem 1]{sontag1995changing} we obtain \eqref{eq:LemmaMultipleChangeSupplyRate_derivative_W}, which concludes the proof. 
   \hfill $\blacksquare$
 
The next lemma is related to the decay rate of the Lyapunov function in Remark \ref{decayRateRemark}. 
 \begin{lem}
 Consider system~\eqref{eq:HybridSystem}-\eqref{eq:jumpMap_i_Esplicit} and suppose Assumptions~\ref{assumption 1}-\ref{ISSassumption} hold. For any $\alpha_U \in \Kinf$ such that $\alpha_U \leq \alpha$, any compact set $\mathcal{M}\subset \mathcal{Q}$ and any $\nu >0$, select $\sigma_i$, $c_i$, $\varepsilon_i$, $d_i$, $b_i$ and $\delta_i$ as in Theorem~\ref{theoremLyapunov} for all $i \in \{1, \dots, N\}$ and define $\overline{d}:= \max\{d_1, \dots, d_N\}$. 
 Select $\alpha_i \in \Kinf$ such that $ \min \left\{\delta_1\alpha_1\left(\frac{s}{\overline{d}N}\right), \dots, \delta_N\alpha_N\left(\frac{s}{\overline{d}N}\right) \right\} \geq \psi_{\mathcal{M}}(s)$ for all $s \geq 0$, where $\psi_{\mathcal{M}} \in \Kinf$ is the modulus of continuity of the function $\alpha_U$ in the compact set $\mathcal{M}$. Then, 
 for any $q \in \mathcal{C} \cap \mathcal{M}$ and any $w \in \mathcal{W}$, 
  \begin{equation}
 	\begin{array}{l}
 		\left\langle \nabla U(q), F(q,w) \right\rangle  
 		\leq -\alpha_U(U(q)) + \nu + \theta(|v|),
 	\end{array}
 	\label{eq:eqLemmaDecayRate}
 \end{equation}
with $U$ defined in \eqref{eq:LyapunovDefinitionU} and $\theta \in \Kinf$ comes from Assumption~\ref{ISSassumption}. Moreover, \eqref{eq:eqLemmaDecayRate} holds globally, i.e., for any $q\in \mathcal{C}$ and $w\in \mathcal{W}$,  when $\alpha \in \Kinf$ is uniformly continuous or when $\alpha \in \Kinf$ is subadditive, i.e. $\alpha(s_1) + \alpha(s_2) \geq \alpha(s_1 + s_2)$, for all $s_1, s_2 \geq 0$ and $\alpha_i \in \Kinf$ with $i \in \{1, \dots, N\}$ are selected such that $ \alpha_i\left(\frac{s}{\overline{d}N}\right) \geq \frac{\alpha(s)}{\delta_i}$ for all $s \geq 0$. 
\hfill	$\Box$

\label{decayRateLemma} 
\end{lem}

 \noindent\textbf{Proof:}
%
 We first show that we can ensure any decay rate $\alpha_U$ on flows for $U$ along solutions to \eqref{eq:HybridSystem}-\eqref{eq:jumpMap_i_Esplicit} with $\alpha_U \in \Kinf$ and $\alpha_U \leq \alpha$ on any given compact set by suitably selecting $\alpha_i$ in \eqref{eq: etaEquation}, for all $i \in \{1, \dots, N\}$. 
 
Let $\mathcal{M} \subset \mathcal{Q}$ be a compact set, $q\in \mathcal{C}\cap\mathcal{M}$ and $w \in \mathcal{W}$, 
 from \eqref{eq:stabilityTheoremFlow} and by using \cite[Lemma 4]{wang2019periodic} 
 we obtain
\begin{equation}
	\begin{array}{l}
		\left\langle \nabla U(q), F(q,w) \right\rangle  \\[.5em]
		\qquad \leq -\alpha (V(x,z))- \alpha_{\eta}\Big(\sum\limits_{i = 1}^{N}d_i\eta_i\Big) + \nu + \theta(|v|), \\
	\end{array}
	\label{eq:eqRemarkNonlinearDecayRate3}
\end{equation}
where $ \alpha_{\eta}(s):= \min\left\{\delta_1{\alpha}_1\left(\frac{s}{\overline{d}N}\right), \dots, \delta_N{\alpha}_N\left(\frac{s}{\overline{d}N}\right)\right\} \in \Kinf$, with $\overline{d}:= \max\{d_1, \dots, d_N\}$. 
 Take any $\alpha_U \in \Kinf$ such that $\alpha_U \leq \alpha$ on $\mathcal{M}$. From the Heine-Canton theorem, we have that 
 $\alpha_U$ is uniformly continuous on $\mathcal{M}$. Applying \cite[Proposition A.2.1]{postoyan2009commande} we have that, for all $q \in \mathcal{M}$, 
$\alpha_U\Big(V(x,z) + \sum\limits_{i = 1}^{N} d_i \eta_i \Big) - \alpha_U\Big(V(x,z)\Big) \leq \psi_{\mathcal{M}}\Big( \sum\limits_{i = 1}^{N} d_i \eta_i\Big),$ 
%
%
where $\psi_{\mathcal{M}} \in \Kinf$ is the modulus of continuity of $\alpha_U$. Selecting $\alpha_i \in \Kinf$, $i \in \{1, \dots, N\}$ such that, for all $s \geq 0$,
		$\alpha_{\eta}(s) = \min \left\{\delta_1\alpha_1\left(\frac{s}{\overline{d}N}\right), \dots, \delta_N\alpha_N\left(\frac{s}{\overline{d}N}\right) \right\} \geq \psi_{\mathcal{M}}(s),$
 we obtain from \eqref{eq:eqRemarkNonlinearDecayRate3},
%
	$	\left\langle \nabla U(q), F(q,w) \right\rangle  
		 \leq -\alpha (V(x,z))- \alpha_{\eta}\Big(\sum\limits_{i = 1}^{N}d_i\eta_i\Big) + \nu + \theta(|v|) 
		 \leq -\alpha (V(x,z))- \alpha_U\Big(V(x,z) + \sum\limits_{i = 1}^{N} d_i \eta_i \Big)  
		 + \alpha_U\Big(V(x,z)\Big) + \nu + \theta(|v|), $
and since $\alpha_U \leq \alpha$, 
%
	$	\left\langle \nabla U(q), F(q,w) \right\rangle  
	 \leq - \alpha_U\Big(V(x,z) + \sum\limits_{i = 1}^{N} d_i \eta_i \Big) + \nu + \theta(|v|) 
		= -\alpha_U(U(q)) + \nu + \theta(|v|). $
 Moreover, when $\alpha \in \Kinf$ is uniformly continuous the result is global for all $\alpha_U \in \Kinf$ such that $\alpha_U \leq \alpha$ and $\alpha_U$ uniformly continuous. This comes directly from the first part of this proof. 
 
We now prove the last part of the lemma, in particular we prove that \eqref{eq:eqLemmaDecayRate} holds globally when $\alpha \in \Kinf$ is subadditive, i.e. $\alpha(s_1) + \alpha(s_2) \geq \alpha(s_1 + s_2)$, for all $s_1, s_2 \geq 0$ and $\alpha_i \in \Kinf$ with $i \in \{1, \dots, N\}$ are selected such that $ \alpha_i\left(\frac{s}{\overline{d}N}\right) \geq \frac{\alpha(s)}{\delta_i}$ for all $s \geq 0$.
 From \eqref{eq:eqRemarkNonlinearDecayRate3} we have 
 		$\left\langle \nabla U(q), F(q,w) \right\rangle 
 		\leq -\alpha (V(x,z))- \alpha_{\eta}\Big(\sum\limits_{i = 1}^{N}d_i\eta_i\Big) + \nu + \theta(|v|), 
 		 \leq -\alpha (V(x,z)) - \alpha\Big(\sum\limits_{i = 1}^{N}d_i\eta_i\Big) + \nu + \theta(|v|),$
 where the last inequality comes from 
		$\alpha_{\eta}(s) = 
		\min \left\{\delta_1\alpha_1\left(\frac{s}{\overline{d}N}\right), \dots, \alpha_N\left(\frac{s}{\overline{d}N}\right) \right\} 
		\geq \alpha(s),$ for all $s \geq 0.$
Since $\alpha$ is subadditive, we obtain
		$\left\langle \nabla U(q), F(q,w) \right\rangle  
		 \leq - \alpha\Big(V(x,z)+ \sum\limits_{i = 1}^{N}d_i\eta_i\Big) + \nu + \theta(|v|)
		 = -\alpha(U(q)) + \nu + \theta(|v|)$
 and since $\alpha_U \leq \alpha$, 
	$\left\langle \nabla U(q), F(q,w) \right\rangle   \leq - \alpha_U(U(q)) + \nu + \theta(|v|). $ 
 \hfill $\blacksquare$

\bibliography{bibliography.bib}

\end{document}